\documentclass[journal, twoside, onecolumn, draftclsnofoot,12pt]{IEEEtran}

\usepackage{multirow}  \usepackage{cite}
\usepackage{graphicx,subfigure}\usepackage{amsmath} \usepackage{amssymb} \usepackage{amsthm}\usepackage{color,array}
\usepackage{etex,etoolbox} \usepackage{float}
\usepackage{color}\usepackage{graphicx}
\usepackage{caption}\usepackage{xcolor}
\usepackage{colortbl}
\usepackage{soul}
\usepackage{algorithm}
\usepackage{soul,color,bm}
\usepackage{setspace}

\usepackage{algorithmic}   \usepackage{bbm}
\usepackage[T1]{fontenc}

\makeatletter
\providecommand{\@fourthoffour}[4]{#4}
\def\fixstatement#1{%
  \AtEndEnvironment{#1}{%
    \xdef\pat@label{\expandafter\expandafter\expandafter
      \@fourthoffour\csname#1\endcsname\space\@currentlabel}}}

\globtoksblk\prooftoks{1000}
\newcounter{proofcount}
\stepcounter{proofcount}
\long\def\proofatend#1\endproofatend{%
  \edef\next{ \Alph{proofcount}. Proof of \pat@label \noexpand\begin{proof}[Proof]}%
  \toks\numexpr\prooftoks+\value{proofcount}\relax=\expandafter{\next#1\end{proof}}
  \stepcounter{proofcount}}

\def\printproofs{%
  \count@=\z@
  \loop
    \the\toks\numexpr\prooftoks+\count@\relax
    \ifnum\count@<\value{proofcount}%
    \advance\count@\@ne
  \repeat}

\makeatother

\fixstatement{thm}
\fixstatement{lem}

\begin{document}


\title{\huge{Deep Learning-based Resource Allocation \\ For Device-to-Device Communication}}
\author{Woongsup Lee and Robert Schober\thanks{W. Lee is with the Department of Information and Communication Engineering, 
Institute of Marine Industry, Gyeongsang National University, Republic of Korea (email: wslee@gnu.ac.kr). R. Schober is with the Institute for Digital Communication at the Friedrich-Alexander-University Erlangen-N\"urnberg, Erlangen, Germany  (email: robert.schober@fau.de).}}

\maketitle



\vspace{-4mm}

\begin{abstract}

In this paper, a deep learning (DL) framework for the optimization of the resource allocation in 
multi-channel cellular systems with device-to-device (D2D) communication is proposed. Thereby, 
the channel assignment and discrete transmit power levels of the D2D users, which are both integer 
variables, are optimized to maximize the overall spectral efficiency 
whilst maintaining the quality-of-service (QoS) of the cellular users. 
Depending on the availability of channel state information (CSI), two different configurations are 
considered, namely 1) centralized operation with full CSI and 2) distributed operation with partial CSI, 
where in the latter case, the CSI is encoded according to the capacity of the feedback 
channel. Instead of solving the resulting resource allocation problem 
for each channel realization, a DL framework is proposed, where 
the optimal resource allocation strategy for arbitrary channel conditions is approximated 
by deep neural network (DNN) models. Furthermore, we propose a
new training strategy that combines supervised and unsupervised learning 
methods and a local CSI sharing strategy to achieve near-optimal performance 
while enforcing the QoS constraints of the
cellular users and efficiently handling the integer optimization variables based on a
few ground-truth labels. Our simulation results confirm that 
near-optimal performance can be attained with low computation time, 
which underlines the real-time capability of the proposed scheme. Moreover, our
results show that not only the resource allocation strategy but also
the CSI encoding strategy can be efficiently determined using a DNN.
Furthermore, we show that the proposed DL framework can be easily extended  
to communications systems with different design objectives.

\end{abstract}
\vspace{-2mm}

\begin{IEEEkeywords}\vspace{-2mm}
Multi-channel wireless communications systems, interference channel, D2D transmission, deep learning, distributed operation, resource allocation, CSI sharing.
\end{IEEEkeywords}

\vspace{-2mm}
\section{Introduction} \vspace{-2mm}

In wireless communication systems (WCS), efficient resource allocation is important
to achieve high system performance. Resource allocation for WCS has been studied 
for several decades for different types of systems including cognitive radio (CR) \cite{Ng2015, ElTanab2016, Lee2018c}, 
device-to-device (D2D)  \cite{Yu2011, Jiang2016, Lee2019b, Fodor2011},
vehicle-to-everything (V2X)  \cite{Pyun2016, Gao2019},
and non-orthogonal multiple access (NOMA) \cite{Sun2016a, Li2016} communication systems.

Existing resource allocation schemes typically aim to maximize spectral 
efficiency (SE) \cite{Lee2018c, Pyun2016, Sun2016a} or energy efficiency 
(EE) \cite{Jiang2016, Lee2019b} or to minimize the total transmit 
power \cite{Ng2015, Li2016} under different constraints, e.g., quality-of-service (QoS) 
\cite{Li2016, Fodor2011} and 
interference \cite{Lee2019b, Ng2015, ElTanab2016} constraints. 
The resource allocation in WCS becomes very challenging
when multiple users transmit data simultaneously over the same 
radio resource, e.g., in D2D WCS, because of the 
resulting multi-user interference. In this case, resource allocation 
algorithm design leads to a nonconvex optimization problem which 
is generally difficult to solve in an efficient manner due to its NP-hardness.

In the literature, most existing resource allocation schemes \cite{Shi2011, Lee2019b, Fodor2011} 
involve iterative algorithms which limits their applicability in practice where real-time operation 
is essential \cite{Sun2018, Lee2018}. In particular, iterative methods may take a long 
time to converge when the number of users or resource allocation variables is large.  
Moreover, such analytical solutions for resource allocation are taylor-made for a 
specific problem and a completely new solution has to
be developed if the problem formulation changes only slightly, e.g.,
a different objective function or additional constraints are considered, 
which further limits the applicability of this approach. Deep learning (DL)-based schemes can overcome 
the aforementioned shortcomings of current 
resource allocation designs.

\vspace{-2mm}
\subsection{Application of DL in WCS}\vspace{-1mm}

DL, which is based on deep neural 
networks (DNNs), has recently seen a surge in interest in different application domains, 
mainly due to its remarkable advantages over conventional methods. 
For example, image classification based on DL has even surpassed 
human-level performance \cite{LeCun2015}. In DL, a DNN model
is trained to minimize the adopted loss function based on a large
amount of data and the back propagation algorithm without 
relying on an explicit
hand-crafted mathematical formulation of the system model. Although DL 
has been extensively studied in the fields of computer vision and natural 
language processing, only recently, it has been seriously considered for 
the design of WCS \cite{Wang2017} regarding problems such as channel estimation, 
data detection, and classification of wireless signals.

Channel estimation and data detection based on DL 
were studied for orthogonal frequency-division multiplexing (OFDM) \cite{Ye2018}, 
filter bank multicarrier modulation \cite{Cheng2019}, 
multiple-input and multiple-output (MIMO) \cite{Samuel2017}, 
millimeter-wave (mmWave) MIMO \cite{He2018}, and 
molecular communication \cite{Farsad2018} systems. Moreover, 
the authors of \cite{Klautau2018} studied DL-based channel estimation 
and detection for MIMO systems with 1-bit analog-to-digital converters.
Furthermore, in \cite{Wen2018}, a DL-based channel state information (CSI) feedback scheme for 
massive MIMO systems was proposed.

Regarding the classification of wireless signals, it was shown
in \cite{Shea2016} and \cite{OShea2016a} that the modulation format
and the type of data traffic can be determined accurately with DNNs. DL
was also applied for cooperative spectrum sensing in CR systems 
\cite{Lee2019}, where a DNN-based cooperative spectrum sensing
strategy was proposed. Moreover, DL was exploited for
the design of encoders and decoders using a DNN-based autoencoder 
for optical wireless communications \cite{Lee2019c} and sparse code 
multiple access (SCMA) \cite{kim2018}.
Finally, the practical relevance and viability of DL-based
WCS design were verified by implementation with 
software-defined radios \cite{Shea2018, Doerner2018}
and corresponding digital circuits using hardware description language \cite{Kim2019}.

\vspace{-2mm}
\subsection{DL-based Resource Allocation}\vspace{-1mm}

Given that DNNs can approximate an arbitrary continuous function, 
allowing them to mimic the behavior of highly nonlinear and complex systems 
and making them universal approximators \cite{Hornik1989}, they can also 
be applied for resource allocation in WCS. More specifically, DNNs can be
trained based on the gradient of a loss function, i.e., by back propagation, 
to emulate the function which characterizes the optimal solution of
a resource allocation problem without having to solve a complicated 
optimization problem explicitly \cite{Lee2018}. Accordingly, a near-optimal 
resource allocation strategy can be found by feeding appropriate input data, e.g., the 
channel gains of the system, to the trained DNN.

DL-based resource allocation is beneficial compared to conventional approaches 
in view of its high adaptability, high flexibility, and low complexity. First, 
DL-based resource allocation does not rely on a theoretical channel model and 
can adapt its operation to the actual environment, leading to a higher performance 
in practice. 
Second, DL-based resource allocation is flexible
because the same DNN structure can be used to achieve different design goals 
by changing the loss function \cite{Lee2018}. Finally, 
the computation time\footnote{The training of the DNN, which 
can be time consuming, can be performed offline before the 
DNN is actually used. Hence, the potentially long training time
is not an obstacle to real-time operation of DNN-based systems.} 
required by the trained DNN to obtain the resource allocation policy is much lower
than that of conventional resource allocation schemes because 
DNNs perform only simple matrix operations \cite{Sun2018, Tan2018}.

Due to its merits, DL has been applied for resource allocation in WCS recently 
\cite{Sun2018, Lee2018, Kerret2018, Liang2018, Lee2018b, Lee2020, Lee2019a, VanChien2019}.
Specifically, according to the type of training, DL-based resource allocation schemes can be 
divided into two categories, namely 1) supervised learning-based schemes and 
2) unsupervised learning-based schemes.

For resource allocation design based on supervised
learning, the optimal resource allocation strategy for a given
channel realization is provided as training data and the DNN is
trained to regenerate that optimal strategy \cite{Sun2018, Sanguinetti2018}. 
In this case, the DNN is used to reduce the computational complexity required
for obtaining the optimal resource allocation policy by emulating iterative computations. 
In \cite{Sun2018}, a DNN was trained to reconstruct the optimal transmit power control
with respect to the weighted minimum mean square error (WMMSE). 
On-off transmit power control was considered in \cite{Kerret2018}, 
where each device decided whether to transmit or not in a distributed manner 
based on the estimated CSI of the channel to other devices. Moreover, EE maximization 
in a multi-user network was studied
\cite{Zappone2018}. Furthermore, in \cite{Sanguinetti2018}, 
the transmit power for max-min and max-prod power allocation in   
a downlink massive MIMO system was investigated, and the optimal transmit power 
was determined based on the location of the users. In addition, in \cite{VanChien2019},
the transmit power of the pilot and data symbols in a massive MIMO system was 
controlled via DNNs where the user activity in each cell was 
taken into account.

Although resource allocation based on supervised learning was
shown to significantly reduce computational complexity, 
it requires the collection of label data for the training of the DNNs. 
To this end, the optimal resource allocation strategy has 
to be obtained for a large number of channel realization 
which can be challenging for complex communications systems with large numbers of 
users. Moreover, this approach is not flexible because completely new 
label data has to be generated when the system model changes, e.g., when
a new constraint is added.

On the other hand, for unsupervised
learning-based resource allocation schemes, the data is not labeled and the DNNs 
autonomously determine the optimal resource allocation strategy 
based on the input samples, e.g., the channel gains. Accordingly, the collection of 
training data is much easier compared to supervised learning-based schemes. 
In \cite{Lee2018}, the transmit power was optimized for maximization of 
the SE and EE using convolutional neural networks (CNNs) and 
unsupervised learning. Specifically, the SE and EE were directly used as loss functions 
for training and the DNNs were trained  
without label data for the optimal power allocation. However, only a constraint on 
the total transmit power was considered and more general constraints, 
e.g., QoS requirements, were not taken into account. In 
\cite{Liang2018} and \cite{Lee2019a}, QoS constraints were considered in addition to 
a total transmit power constraint by including the QoS violation probability in the loss
function. With this approach, the DNN was trained to increase the SE while concurrently
reducing the probability of violating the constraints. Moreover, 
to further improve the performance, an ensemble of DNN models was proposed in 
\cite{Liang2018}. However, both \cite{Liang2018} and \cite{Lee2019a} 
do not consider multi-channel systems which limits the 
applicability of the derived schemes. In \cite{Lee2018b} and \cite{Lee2020}, 
DL-based resource allocation schemes with QoS constraints were
proposed for multi-channel CR and D2D communications, respectively.
Finally, in \cite{Lee2019d}, a generic DL-based resource allocation 
framework to handle non-convex constrained optimization problems
was provided and its distributed implementation was considered.
Most existing proposals for unsupervised learning-based resource allocation 
assume that perfect global CSI is available and the optimization variables are continuous.
However, these assumptions are not realistic because the global CSI has to be collected 
via a feedback channel which has a limited capacity in practice. Furthermore, 
some optimization variables may be integer such as the index of the assigned 
channels, which requires a special training methodology to ensure that the output of the DNN is discrete.
Accordingly, in this work, we propose new 
DL-based resource allocation schemes which address the aforementioned shortcomings of 
previous approaches.

\vspace{-2mm}
\subsection{Contributions and Organization}\vspace{-1mm}

In this paper, we propose DL-based resource allocation strategies for 
multi-channel cellular systems with interfering users. In particular, 
we consider a D2D communication-enabled cellular system where 
multiple users transmit data over the same channel simultaneously, 
causing co-channel interference. The main contributions of this paper 
can be summarized as follows.

\begin{enumerate}
\item We propose DNN-aided resource allocation strategies for multi-channel D2D enabled cellular 
systems where the channel indices and the discrete transmit power levels 
of the users are optimally selected 
such that the SE of the D2D users is maximized while  
a minimum data rate is guaranteed for the legacy cellular users. Novel DNN models and a
novel training methodology are proposed to efficiently handle the involved discrete 
optimization variables. A hybrid training strategy, which combines supervised 
and unsupervised training, is adopted such that the computation time for the 
training is low. Although we mainly focus on the maximization of the 
SE in D2D enabled cellular systems, the proposed DL framework can be extended with minor 
modifications to other design objectives  
because the DNN structure does not rely on a specific system model.

\item Both centralized and distributed resource allocation schemes are developed. In particular, for 
the distributed scheme, the optimal sharing strategy for the local CSI, i.e., 
the optimal encoding strategy of the local CSI sent from individual users to the BS 
and the optimal encoding strategy  of the collected local CSI sent from the BS to the users, 
is determined via DL. Thereby, the limited capacity of 
the feedback channel is taken into account.

\item The performance of the proposed schemes is evaluated via computer 
simulations under various conditions. Our simulation results confirm 
that the proposed scheme can achieve near-optimal performance with low 
computation time. We also show that the proposed hybrid training 
strategy is beneficial for reducing the training time of DNNs and the optimal
encoding of the local CSI can be efficiently learned with the proposed
DL-based resource allocation schemes.
Furthermore, we show that the proposed DNN model can be 
easily modified to address alternative design objectives, e.g., the maximization
of the EE.

\end{enumerate}

The remainder of this paper is organized as follows. In Section II,
we introduce the considered system model and problem formulation. 
The proposed DNN structure for resource allocation and the proposed 
DNN training methodology are provided
in Sections III and IV, respectively. Simulation results are given
in Section V, and conclusions are drawn in Section VI.

\textit{Notations: } We use upper-case boldface letters, lower-case boldface letters, 
and normal letters for matrices, vectors, and scalars, respectively. $\{0, 1 \}^{m}$ denotes 
a set of binary values whose cardinality is $m$, and $\mathbb{E}_{h}[\cdot]$ 
denotes the expectation with respect to random variable $h$.

\section{System Model and Problem Formulation}\vspace{-1mm}

We first present the adopted channel model. Then, the problem statement is provided, and the considered
resource allocation problems are formulated.

\vspace{-1mm}
\subsection{Channel Model}\vspace{-1mm}

\begin{figure}[t]
	\centerline{\includegraphics[width=12.0cm]{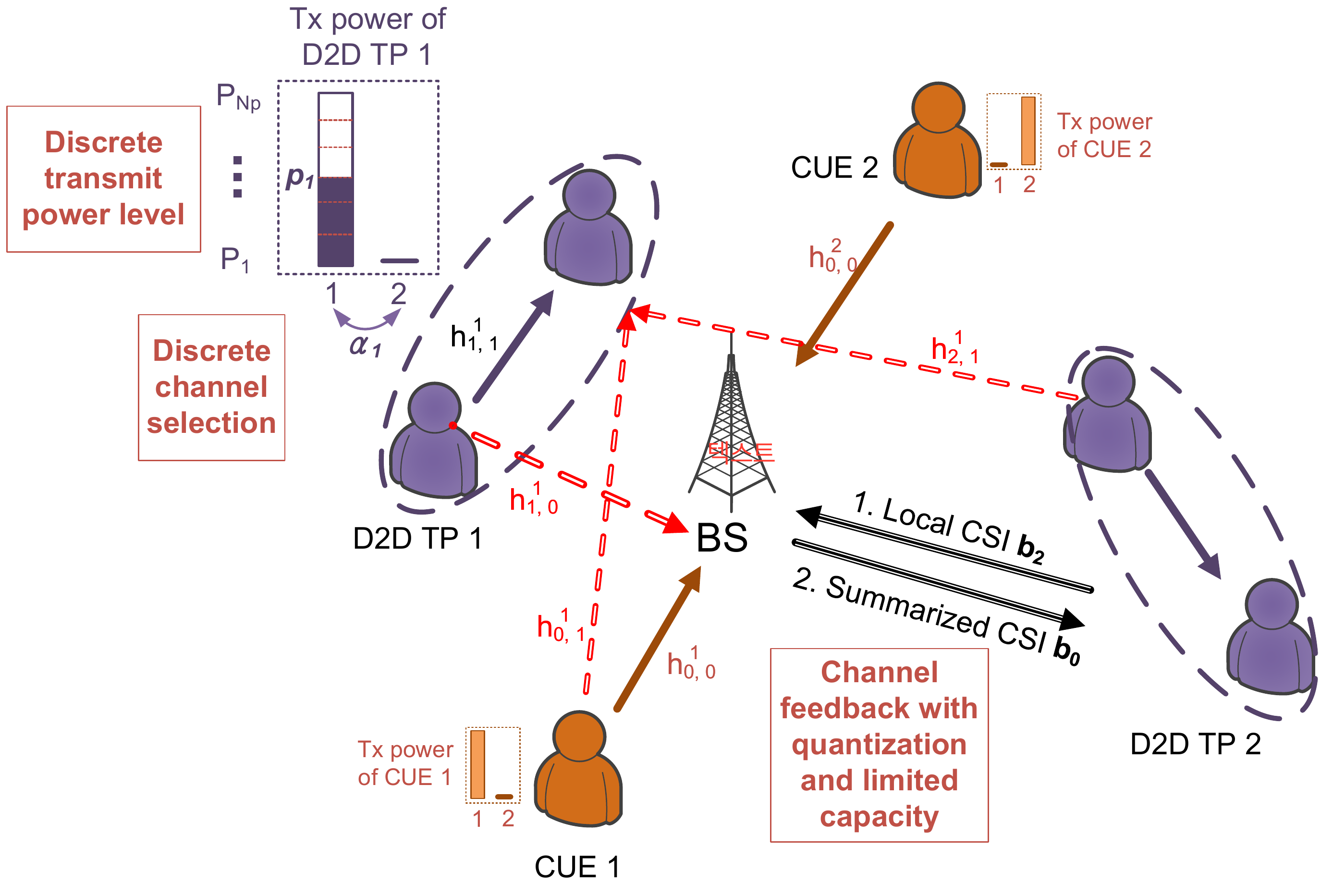}}\vspace{-1mm}
	\caption{System model with two D2D TPs and two channels ($N = K = 2$). For clarity of illustration, not all links are shown in the figure.}
	\label{fig_system_model} \vspace{-3mm}
\end{figure}

In the considered system, there are $K$ channels, where 
$\mathbb{K} = \{1, \cdots , K \}$ denotes the set of channels,
and in each channel, one cellular user equipment (CUE) transmits 
data in the uplink to the BS. The number of D2D transmit pairs 
(TPs) is set to $N$ where each TP consists of one transmitter and one 
receiver, and the set of D2D TPs is denoted by $\mathbb{I} = \{1, \cdots , N \}$. 
Moreover, we assume that all D2D TPs and CUEs are randomly distributed over an area 
$\mathcal{S}$ and are equipped with a single antenna. 
The considered system model for $K$ = $N$ = 2 is depicted in 
Fig. \ref{fig_system_model}.

The channel gain of the $k$-th channel, which comprises both the 
distance-dependent channel gain (i.e., path loss) and multipath fading, 
is denoted as $h^{k}_{i, j}$, where $i\neq0$ and
$j\neq0$  are the indices of the D2D transmitters and receivers, respectively, and
$i=0$ and $j=0$ are the indices of the CUEs and the BS, respectively. 
For example, $h^{1}_{0, 1}$ is the channel gain between the CUE and the 
receiver of the first D2D TP in the first channel, see Fig. \ref{fig_system_model}. 
Furthermore, $\bm{h}_{i} = [h^{1}_{i, 0} \cdots h^{K}_{i, N}]$ 
denotes the vector of channel gains that D2D TP $i$ can acquire,
i.e., its local CSI, and $\bm{H}$ is the matrix containing all $h^{k}_{i, j}$,
i.e., the global CSI. In addition, sets $\mathbbm{H}$ and 
$\mathbbm{h}$ contain all possible realization of 
$\bm{H}$ and $\bm{h}_{i}$, $\forall i \in \mathbb{I}$, respectively.

We assume that the D2D TPs cannot utilize multiple channels 
simultaneously. Specifically, $a^{k}_{i} \in \{0, 1\}$ indicates whether D2D TP $i$
uses the $k$-th channel, i.e., $a^{k}_{i}=1$ if D2D TP $i$ 
transmits data over channel $k$ and  $a^{k}_{i}=0$ otherwise. Given 
that each D2D TP can utilize only one channel at a time, 
$\sum_{k \in \mathbb{K}} a^{k}_{i} \leq 1, \forall i \in \mathcal{I}$. 
Moreover, we define $\bm{a}_{i} = [a^{1}_{i}, \cdots,  a^{K}_{i}]$ as
the vector of channel usage indicators for D2D TP $i$.
Unlike in \cite{Lee2018b} and \cite{Lee2020}, where users can utilize multiple 
channels simultaneously, the consideration of single channel usage
results in an integer programming problem, which is difficult 
to solve efficiently.

The transmit power of D2D TP $i$ is denoted as $p_{i}$. 
We assume that the transmit power of the users
is divided into $\textrm{N}_{\textrm{P}}$ discrete levels such that $p_{i} \in \mathcal{P}
= \{\textrm{P}_{1}, \cdots, \textrm{P}_{\textrm{N}_{\textrm{P}}}\}$, 
where the elements of $\mathcal{P}$ are organized in ascending order 
such that $\textrm{P}_{1} = 0$ and $\textrm{P}_{\textrm{N}_{\textrm{P}}} = \textrm{P}_{\textbf{\textrm{M}}}$. 
Herein, $\textrm{P}_{\textbf{\textrm{M}}}$ is the maximum transmit power of each user.
The consideration of discrete power levels, which are used in many communication 
standards, e.g., 3GPP LTE and IS-95 \cite{Nguyen2015, Liu2015}, 
is more realistic than assuming continuous power levels. Moreover, 
discrete power levels are beneficial when transmitters have 
limited capabilities due to hardware constraints, e.g., in machine-type 
communications \cite{Suciu2017}. However, optimization problems involving 
discrete power levels are more difficult to solve than problems 
with continuous power levels because of the resulting integer programming. 
Furthermore, we assume that the transmit power of the CUEs is 
$\textrm{P}_{\textrm{C}}$. In addition, we denote the bandwidth of each channel 
and the noise power spectral density by $W$ and $N_{0}$, respectively.

\subsection{Problem Statement}

In this paper, the objective of resource allocation is to maximize\footnote{We note that
the proposed approach is general and allows the use of other objective functions, e.g., 
the maximization of the EE, and the incorporation of additional constraints. 
In Section V, we show that the proposed DNN structure can also be employed to 
maximize the sum EE of the D2D TPs.} 
the sum SE of the D2D TPs while guaranteeing that the SE of the 
CUEs does not fall below a certain threshold, $\textrm{SE}_{\textrm{thr}}$, 
which constitutes a QoS requirement. To this end, the selection 
of the channel, $\bm{a}_{i}$, and the transmit power level, $p_{i}$, have to 
be optimized according to the current channel conditions. 

According to the universal approximation theorem \cite{Hornik1989, Sun2018, Lee2019d}, 
DNNs can approximate arbitrary functions. Therefore, instead of solving 
the optimization problem for resource allocation for each channel realization, 
functions that approximate the optimal resource allocation strategy 
for arbitrary channel conditions can be realized via DNNs. 
Then, the optimal resource allocation is found by feeding the current 
channel gains to the DNN instead of solving the 
optimization problem independently for each channel 
realization \cite{Lee2018, Lee2018b, Lee2020}.

Regarding the availability of CSI, we consider two cases, namely 
full CSI and partial CSI, which lead to centralized and distributed resource 
allocation policies, respectively. In the former case, 
the BS is able to collect the complete CSI, $\bm{H}$, such that
the resource allocation for each D2D TP
can be determined in a centralized manner. In the latter case, the CSI
is shared via CSI feedback channels with limited capacity
such that the resource allocation has to be determined 
in a distributed manner based on partial CSI. Moreover, 
for the case of partial CSI, the local CSI has to be encoded, 
which has to be taken into account for the optimization. In the following two 
subsections, we provide the detailed problem formulations
for centralized and distributed resource allocation, respectively.

\subsection{Centralized Resource Allocation}

For centralized resource allocation, full CSI $\bm{H}$ is assumed to be available and
our objective is to find the optimal resource allocation policy for this case. 
Let $f_{c, a}^{i, k}(\cdot)$ and $f_{c, p}^{i}(\cdot)$ denote the functions 
that approximate the optimal channel selection and transmit
power level of D2D TP $i$, such that $a^{k}_{i}  = f_{c, a}^{i, k}(\bm{H})$ 
and $p_{i} = f_{c, p}^{i}(\bm{H})$. In the considered D2D system model, the receiver of D2D TP 
$i \in \mathbb{I}$ receives data from the transmitter of D2D TP $i \in \mathbb{I}$ 
and interference from the transmitters of the other D2D TPs $l \in \mathbb{I}\setminus \{i\}$ and the CUE. 
Accordingly, the achievable SE of D2D TP $i$ in the $k$-th channel, $\textrm{SE}^{k}_{i}$, 
with resource allocation policy $f_{c, a}^{i, k}$ and $f_{c, p}^{i}$
is given as follows:
\begin{equation}
\begin{array}{lll}
\textrm{SE}^{k}_{i} 
= \log_{2} \left(1 + \dfrac{  h^{k}_{i,i} f_{c, a}^{i, k}(\bm{H})f_{c, p}^{i}(\bm{H}) }{N_{0}W + \sum \limits_{l\in\mathbb{I} \setminus \{i\} } h^{k}_{i, l} f_{c, a}^{l, k}(\bm{H}) f_{c, p}^{l}(\bm{H}) +  h^{k}_{i, 0} \textrm{P}_{\textrm{C}}  } \right),
 \label{SE_eq1}
\end{array}
\end{equation}
where $h^{k}_{i, 0} \textrm{P}_{\textrm{C}}$ corresponds to the interference from the CUE using channel $k$.

Similarly, the SE of the CUE using channel $k$, which we denote by $\textrm{SE}^{k}_{0}$, can 
be written as follows:
\begin{equation}
\begin{array}{lll}
\textrm{SE}^{k}_{0} 
= \log_{2} \left(1+\dfrac{  h^{k}_{0,0} \textrm{P}_{\textrm{C}}  } {N_{0}W + \sum \limits_{l\in\mathbb{I} } h^{k}_{0, l} f_{c, a}^{l, k}(\bm{H})f_{c, p}^{l}(\bm{H})  }\right).
 \label{SE_eq2}
\end{array}
\end{equation}

Then, the functions $f_{c, a}^{i, k}(\cdot)$ and $f_{c, p}^{i}(\cdot)$, which maximize 
the overall SE of the D2D TPs subject to a minimum required SE of the CUE, $\textrm{SE}_{\textrm{thr}}$,
can be found by solving the following optimization problem:\vspace{-1mm}
\begin{equation}
\begin{array}{lll}
\underset{f_{c, a}^{i, k}, f_{c, p}^{i}} {\operatorname{maximize}} &&  \mathbb{E}_{\bm{H}} \left[ \sum \limits_{i \in \mathbb{I} } \sum \limits_{k \in \mathbb{K} } \textrm{SE}^{k}_{i} \right]\\
~~~\textrm{s.t.} && f_{c, a}^{i, k}(\bm{H}) \in \{0, 1\} ~~~~~~~~\forall  i \in \mathbb{I},~\forall k \in \mathbb{K},~\forall \bm{H} \in \mathbbm{H}, \\

~~~&&  \sum_{k \in \mathbb{K}} f_{c, a}^{i, k}(\bm{H})
\leq 1 ~~~~~~\forall i \in \mathbb{I},~\forall \bm{H} \in \mathbbm{H},\\

~~~&&   f_{c, p}^{i}(\bm{H})\in \mathcal{P} ~~~~~~~~~~~~~ \forall  i \in \mathbb{I},~\forall \bm{H} \in \mathbbm{H},\\

~~~&& \textrm{SE}^{k}_{0}  \geq \textrm{SE}_{\textrm{thr}} ~~~~~~~~~~~~~~ \forall  k \in \mathbb{K},~\forall \bm{H} \in \mathbbm{H}.
  \label{prob_form_1}\vspace{-1mm}
\end{array}
\end{equation}

In (\ref{prob_form_1}), the expected SE is maximized instead of the instantaneous SE since 
functions, $f_{c, a}^{i, k}(\cdot)$ and $f_{c, p}^{i}(\cdot)$, which are optimal for 
any given channel condition, $\bm{H}$, are designed. 


\vspace{-1mm}
\subsection{Distributed Resource Allocation}\vspace{-1mm}

For distributed resource allocation, the local CSI is shared among 
the D2D users, and then each D2D user determines its own resource allocation 
in a distributed manner based on the collected partial CSI. Accordingly, 
the sharing of local CSI needs to be optimized along with the resource
allocation. Herein, considering how CSI is acquired  
in existing cellular communication standards, 
e.g., 3GPP LTE \cite{Militano2015, Cottatellucci2016}, 
we assume that each D2D user sends the local CSI to the BS first, and then 
the BS broadcasts the collected CSI to the D2D users. Although the direct sharing of 
local CSI among users has been considered in the literature \cite{Kerret2018, Lee2019d}, 
it can cause a huge signaling overhead and some D2D users may not be 
able to share their local CSI because of the large distances 
between them.

For distributed resource allocation, we assume that the capacity of the CSI 
feedback channel is limited such that each D2D TP compresses its local CSI 
to $B_{F}$ bits and sends the compressed local CSI to the BS, see 
Fig. \ref{fig_system_model}. The local CSI reported by D2D TP $i$ 
to the BS is denoted as $\bm{b}_{i} \in \{0, 1\}^{B_{F}}: \forall i \in \mathcal{I}$. 
The encoding of the local CSI is modeled via function $z_{i}(\cdot)$ for
given local CSI $\bm{h}_{i}$ such that $\bm{b}_{i} = z_{i}(\bm{h}_{i} )$.
Then, the BS accumulates the local CSI from all D2D TPs,  
$\bm{b}_{i}$, and broadcasts the collected local 
CSI and its own CSI, $\bm{h}_{0}$, to all D2D TPs, 
where the broadcast CSI is compressed into $B_{B}$ bits by the BS. 
The CSI sent from the BS to all D2D TPs is 
denoted by $\bm{b}_{0} \in \{0, 1\}^{B_{B}}$ and 
determined by function $z_{0}(\cdot)$ for given local
CSI of the channel of the CUE at the BS, $\bm{h}_{0}$, and encoded local CSI of
the D2D TPs, $\bm{b}_{i}$, such that $\bm{b}_{0} = z_{0}(\bm{h}_{0}, \bm{b}_{1}, 
\cdots, \bm{b}_{N})$. Note that the total signaling overhead for distributed
resource allocation is $N \cdot B_{F} + B_{B}$ bits because
each D2D TP  reports its own local CSI using $B_{F}$ bits and the BS 
broadcasts the accumulated local CSI to the D2D TPs using $B_{B}$ bits.

Accordingly, for distributed resource allocation, the optimal 
encoding strategies for local CSI sharing, i.e., $z_{i}(\cdot )$ 
and $z_{0}(\cdot)$, have to be determined. In addition, 
the optimal resource allocation strategy of the D2D TPs has to
be obtained based on local CSI, $\bm{h}_{i}$, and the 
combined CSI sent by the BS, $\bm{b}_{0}$. Let $f_{d, a}^{i, k}(\cdot)$ 
and $f_{d, p}^{i}(\cdot)$ be the functions that model the  
channel selection and the transmit power level of D2D TP $i$ 
for partial CSI, such that $f_{d, a}^{i, k}(\bm{h}_{i}, \bm{b}_{0}) = a^{k}_{i}$ and 
$f_{d, p}^{i}(\bm{h}_{i}, \bm{b}_{0}) = p_{i}$.

Then, the following optimization problem must be solved for distributed resource allocation:\vspace{-1mm}
\begin{equation}
\begin{array}{lll}
\underset{z_{0}, z_{i}, f_{d, a}^{i, k}, f_{d, p}^{i}} {\operatorname{maximize}} &&  \mathbb{E}_{\bm{H}} \left[ \sum \limits_{i \in \mathbb{I} } \sum \limits_{k \in \mathbb{K} } \tilde{\textrm{SE}}^{k}_{i} \right]\\
~~~\textrm{s.t.} && f_{d, a}^{i, k}(\bm{h}_{i}, \bm{b}_{0}) \in \{0, 1\} ~~~~~~~~~~~~~~~~~~~~~~~~\forall  i \in \mathbb{I},~\forall k \in \mathbb{K},~\forall \bm{h}_{i} \in \mathbbm{h},~\forall \bm{b}_{0},\\

~~~&&  \sum_{k \in \mathbb{K}} f_{d, a}^{i, k}(\bm{h}_{i}, \bm{b}_{0})
\leq 1 ~~~~~~~~~~~~~~~~~~~~~~\forall i \in \mathbb{I},~\forall \bm{h}_{i} \in \mathbbm{h},~\forall \bm{b}_{0},\\

~~~&&   f_{d, p}^{i}(\bm{h}_{i}, \bm{b}_{0})\in \mathcal{P} ~~~~~~~~~~~~~~~~~~~~~~~~~~~~~ \forall  i \in \mathbb{I},~\forall \bm{h}_{i} \in \mathbbm{h},~\forall \bm{b}_{0},\\

~~~&&  \tilde{\textrm{SE}}^{k}_{0} \geq \textrm{SE}_{\textrm{thr}} ~~~~~~~~~~~~~~~~~~~~~~~~~~~~~~~~~~ \forall  k \in \mathbb{K},~\forall \bm{H} \in \mathbbm{H},~\forall \bm{b}_{0},\\

~~~&&  z_{i}(\bm{h}_{i}) \in \{0, 1\}^{B_{F}} ~~~~~~~~~~~~~~~~~~~~~~~~~~~\forall i \in \mathcal{I},~\forall \bm{h}_{i} \in \mathbbm{h},\\

~~~&&  z_{0}(\bm{h}_{0}, z_{1}(\bm{h}_{1}), \cdots, z_{1}(\bm{h}_{N})) \in \{0, 1\}^{B_{B}}~~\forall \bm{H} \in \mathbbm{H},\\

~~~&&  \bm{b}_{0} = z_{0}(\bm{h}_{0}, z_{1}(\bm{h}_{1}), \cdots, z_{1}(\bm{h}_{N}))~~~~~~~~~\forall \bm{H} \in \mathbbm{H},
  \label{prob_form_1_2}\vspace{-1mm}
\end{array}
\end{equation}
where $\tilde{\textrm{SE}}^{k}_{i} $ and $\tilde{\textrm{SE}}^{k}_{0}$ are given as 
follows
\begin{equation}
\begin{array}{lll}
\tilde{\textrm{SE}}^{k}_{i}  = \log_{2} \left(1 + \dfrac{  h^{k}_{i,i} f_{d, a}^{i, k}(\bm{h}_{i}, \bm{b}_{0})f_{d, p}^{i}(\bm{h}_{i}, \bm{b}_{0}) }{N_{0}W + \sum \limits_{l\in\mathbb{I} \setminus \{i\} } h^{k}_{i, l} f_{d, a}^{l, k}(\bm{h}_{i}, \bm{b}_{0}) f_{d, p}^{l}(\bm{h}_{i}, \bm{b}_{0}) +  h^{k}_{i, 0} \textrm{P}_{\textrm{C}}  } \right),\\
\tilde{\textrm{SE}}^{k}_{0}
= \log_{2} \left(1+\dfrac{  h^{k}_{0,0} \textrm{P}_{\textrm{C}}  } {N_{0}W + \sum \limits_{l\in\mathbb{I} } h^{k}_{0, l} f_{d, a}^{l, k}(\bm{h}_{i}, \bm{b}_{0})f_{d, p}^{l}(\bm{h}_{i}, \bm{b}_{0})  }\right).
 \label{SE_eq_1_3}
\end{array}
\end{equation}

Since the resource allocation strategies of all D2D TP need to be jointly
optimized to achieve the common goal of maximizing the sum SE, a centralized 
optimization problem is formulated in (\ref{prob_form_1_2}), i.e., 
the strategies for channel and transmit power level selection for each D2D TP are 
jointly derived. However, each D2D TP can determine its resource allocation strategy
individually solely based on local CSI $\bm{h}_{i}$ and the summarized CSI coming from the 
BS $\bm{b}_{0}$, because $f_{d, a}^{i, k}(\cdot)$ and $f_{d, p}^{i}(\cdot)$ are the functions
of $\bm{h}_{i}$ and $\bm{b}_{0}$.

\section{DNN Structure For Deep Learning Based Resource Allocation}
In this section, we present the proposed DNN structure used to approximate the optimal
resource allocation functions for centralized and distributed resource allocation, i.e.,
$f_{c, a}^{i, k}$, $f_{c, p}^{i}$, $z_{0}$, $z_{i}$, $f_{d, a}^{i, k}$, and $f_{d, p}^{i}$.

\vspace{-2mm}
\subsection{Structure of Basic DNN Module}\vspace{-1mm}

\begin{figure}[t]
	\centerline{\includegraphics[width=10.0cm]{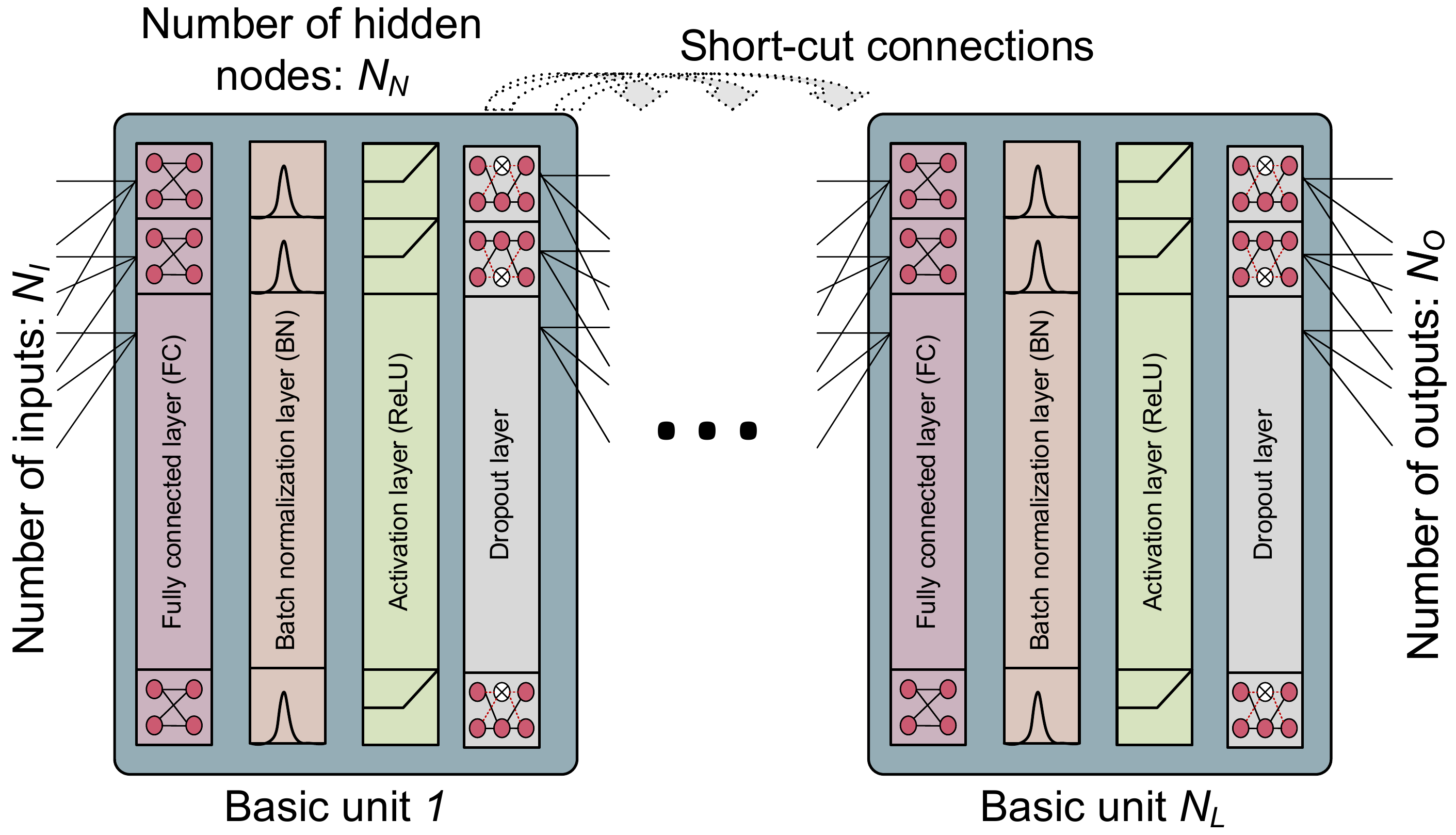}}\vspace{-1mm}
	\caption{Structure of basic DNN module.}
	\label{fig_basic_dnn_structure} \vspace{-8mm}
\end{figure}

In this subsection, we describe the basic DNN module which constitutes the basic building block
of the proposed DNNs. The basic DNN module is composed of multiple basic units
which are in turn composed of one fully connected (FC) layer, one batch normalization (BN) 
layer, one rectified linear unit (ReLU) layer, and one dropout layer, as depicted in 
Fig. \ref{fig_basic_dnn_structure}. These basic DNN 
modules are used to approximate functions, 
$f_{c, a}^{i, k}$, $f_{c, p}^{i}$, $z_{0}$, $z_{i}$, $f_{d, a}^{i, k}$, and $f_{d, p}^{i}$.
 
In the FC layer, the output is determined by the multiplication of weights and the 
addition of a bias to the input \cite{Lee2019a, Goodfellow2016}. Let 
$\textbf{x}_{\textrm{FC}}$, $\textbf{W}_{\textrm{FC}}$, and $\textbf{b}_{\textrm{FC}}$ denote the input vector, weight matrix, and bias vector of the FC layer, respectively. Then, the output is given by $\textbf{W}_{\textrm{FC}} \textbf{x}_{\textrm{FC}} + \textbf{b}_{\textrm{FC}}$.

The output of the FC layer is fed into the 
BN layer which performs mini batch-wise normalization \cite{Goodfellow2016, Ioffe2015b}. 
In general, during the training of DNNs,
multiple input data, i.e., mini batches, are fed into the DNN simultaneously
in order to reduce the computation time via parallel processing. 
In the BN layer, mini batch data is normalized
by its own mean and variance. Then, the normalized data is multiplied 
by weights and biases are added whose values are adjusted during training. 
Mathematically, let $x_{\textrm{BN}}$ be the input of 
the BN layer, then the output of the BN layer is given by
$w_{\textrm{BN}} \frac{x_{\textrm{BN}}-\mu_{\textrm{BN}}}{\sqrt{\sigma^2_{\textrm{BN}}  
+ \epsilon}} + b_{\textrm{BN}}$, where $w_{\textrm{BN}}$ and $b_{\textrm{BN}}$
are the weight and bias of the BN layer, and $\epsilon$ is a small constant
to prevent division by zero. Moreover, $\mu_{\textrm{BN}} = \mathbb{E}_{x_{\textrm{BN}}}[x_{\textrm{BN}}]$ and
$\sigma^2_{\textrm{BN}} = \mathbb{E}_{x_{\textrm{BN}}}[(x_{\textrm{BN}} - \mu_{\textrm{BN}})^{2}]$.
It has been shown that the use of a BN layer is an effective 
means to avoid the overfitting problem 
and the vanishing gradient problem in DNNs by regulating the distribution of the 
inputs of the subsequent layers \cite{Ioffe2015b}.

Then, the output of the BN layer is fed into the ReLU layer 
which introduces nonlinearity to the DNN \cite{Simonyan2014, Lee2019a}. Specifically,
for input vector $\textbf{x}_{\textrm{ReLU}}$, 
the output of the ReLU layer is $\max(\textbf{x}_{\textrm{ReLU}}, 0)$, 
i.e., negative inputs are blocked by the ReLU layer. The use of 
ReLU is also beneficial in view of computation time because its 
output and derivative are very easy to calculate.

Finally, the dropout layer is applied to the output of the ReLU layer.
In the dropout layer, the input data is randomly dropped with a certain 
probability such that some of the input data is blocked from being forwarded 
to the subsequent layer. This drop out is an efficient regularization
technique to avoid the overfitting of DNNs \cite{Lee2018}.
\begin{figure*}[t]
	\centerline{\includegraphics[width=13.5cm]{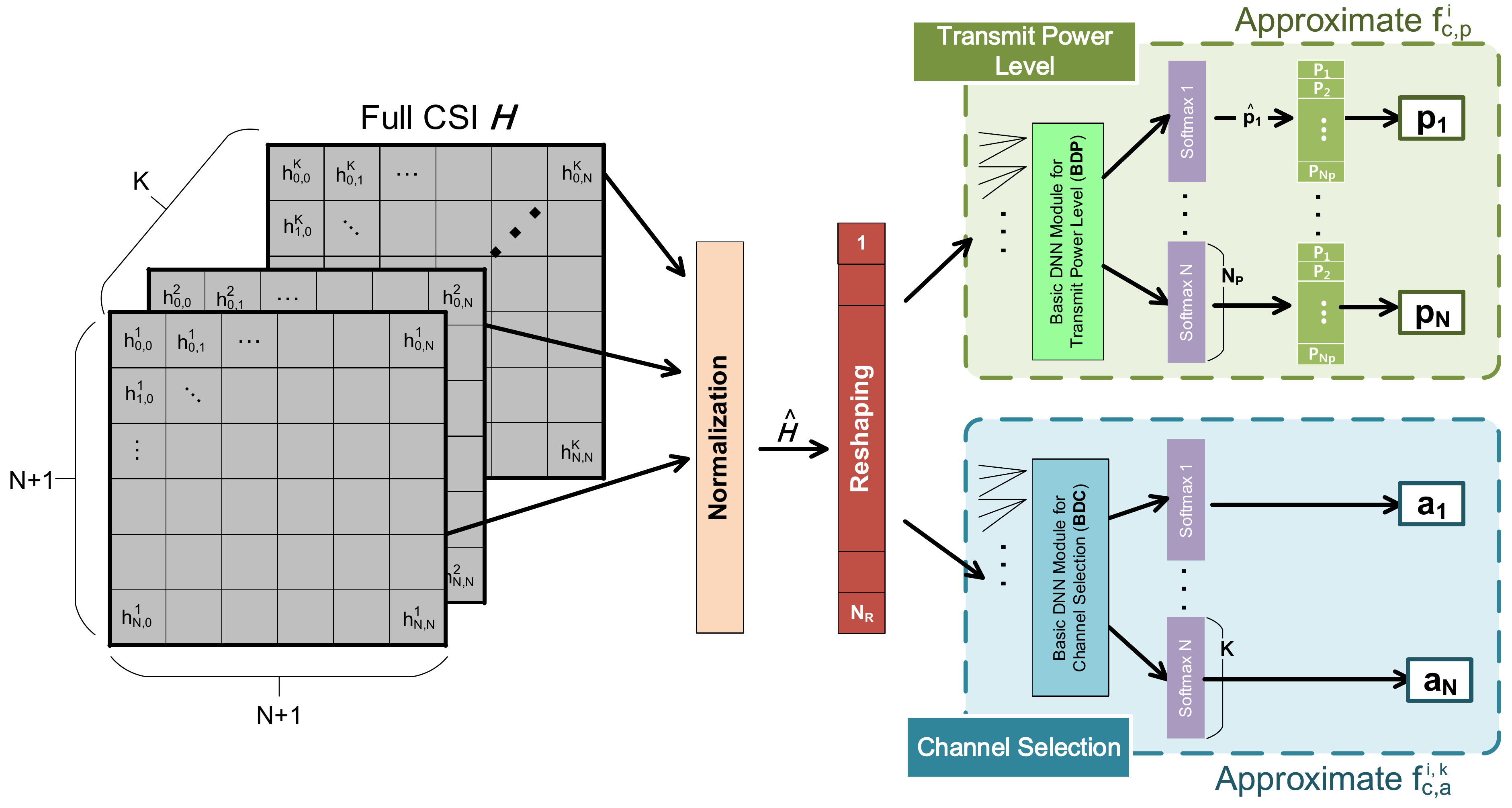}}\vspace{-2mm}
	\caption{Structure of centralized DNN model.}
	\label{fig_full_dnn_structure}\vspace{-6mm}
\end{figure*}

The proposed basic DNN module is composed of $N_{L}$ basic units which are sequentially 
connected, cf. Fig. \ref{fig_basic_dnn_structure}. The numbers of inputs and 
outputs of a basic DNN module are denoted by $N_{I}$ and $N_{O}$, respectively. 
Moreover, the number of hidden nodes of each basic unit, which corresponds to 
the size of the FC layer, is given by $N_{N}$, except for 
the last basic unit, whose number of hidden nodes is equal to
the number of outputs of the basic unit, i.e., $N_{O}$. 
Furthermore, inspired by recent advances in DL, we consider residual 
connections (short-cut connections) in the basic DNN module, such that the output 
of the first basic unit is forwarded to the subsequent units via short-cut 
connections \cite{He2016}. More specifically, the output of the first 
basic unit is added to the output of the BN layer of the subsequent units.
These residual connections can help in the training of the DNN \cite{He2016}.

\vspace{-2mm}
\subsection{Structure of Centralized DNN Model}\vspace{-1mm}

In this subsection, we present the structure of the centralized DNN model which is used to determine the resource allocation based on full CSI, cf. Fig. \ref{fig_full_dnn_structure}. In the centralized DNN model, the complete CSI, $\bm{H}$, is applied as input and the transmit power level, $p_{i}$, and the channel selection, $a^{k}_{i}$, are the outputs. Moreover, the centralized DNN model is implemented at the BS, which collects the CSI from the users and informs the resource allocation to each D2D TP. First, the input is pre-processed by converting it to the dB scale and normalized to have zero mean and unit variance which facilitates proper training of the DNN \cite{Lee2018}. The dB scale is preferable compared to the decimal scale due to the reduced range of possible channel gains. 

Let $\bm{\hat{H}}$ be the matrix of pre-processed channel gains. The elements of 
$\bm{\hat{H}}$, which are denoted by $\hat{h}^{k}_{i, j}$, are given as follows:\vspace{-1mm}
\begin{equation}
\hat{h}^{k}_{i, j} = \dfrac{  \log_{10} (h^{k}_{i, j}) - \mathbb{E} [\log_{10} (h^{k}_{i, j})]} {
\sqrt{\mathbb{E} [(\log_{10} (h^{k}_{i, j})  - \mathbb{E}[\log_{10} (h^{k}_{i, j})])^2]}}.\label{prob_preprocess}\vspace{-1mm}
\end{equation}

After pre-processing of the data, $\bm{\hat{H}}$ is reshaped into a 
one-dimensional vector of length $N_{\textrm{R}} = K(N+1)^2$, 
and then fed into two separate basic DNN modules, which are the basic
DNN module for the transmit power level (BDP) and the basic
DNN module for the channel selection (BDC). The BDP and BDC modules independently 
determine the transmit power level and channel selection,
respectively, i.e., they are used to approximate functions 
$f_{c, p}^{i}$ and $f_{c, a}^{i, k}$.

The number of outputs of the BDP module, which calculates the transmit power 
levels of all D2D TPs, is $N_{\textrm{P}} \cdot N$. The outputs
of the BDP module are divided into $N$ groups, where each group contains $N_{\textrm{P}}$
elements, respectively. Then, each group of outputs is fed into separate 
softmax layer blocks, which perform the softmax operation, such that when the 
input of the $i$-th softmax layer block is $y^j_i$, the $j$-th output of 
the $i$-th softmax layer block becomes $\frac{e^{y^j_i}}{\sum_j e^{y^j_i}}$ \cite{Goodfellow2016}. The output of the $i$-th softmax layer block, which we denote as $\hat{\bm{p}}_{i}$, specifies the probabilities with which each transmit power level is selected, i.e., the $j$-th element of $\hat{\bm{p}}_{i}$ corresponds to the probability that D2D TP $i$ uses transmit power level, $\textrm{P}_{j}$.

The determination of the transmit power level for D2D TP $i$ is different for training and inference. 
During training, the transmit power level of D2D TP $i$ is determined by the multiplication of $\hat{\bm{p}}_{i}$ 
and the discrete transmit power levels. Mathematically,  $p_{i} = \sum \limits_{1 \leq j \leq \textrm{N}_{\textrm{P}}}  
\textrm{P}_{j} \hat{p}^{j}_{i}$, where  $\hat{p}^{j}_{i}$ is the $j$-th element of $\hat{\bm{p}}_{i}$. 
On the other hand, during inference when the D2D TP actually uses the trained DNN to determine its
transmit power, $\sum \limits_{1 \leq j \leq \textrm{N}_{\textrm{P}}}  
\textrm{P}_{j} \hat{p}^{j}_{i}$ cannot be used because the constraint $p_{i} \in \mathcal{P}$ can be violated.
Instead, the index of $\hat{p}^{j}_{i}$ which corresponds to the largest value is employed to select the transmit power level, 
such that $p_{i} = \textrm{P}_{j^*}$ where $j^* = \underset{j}{\operatorname{argmax}} ~\hat{p}^{j}_{i}$.
Note that the $p_{i}$ used during training is differentiable, and hence, back-propagation based training can be used. 
However, the $p_{i}$ used during inference cannot be used for training because $\operatorname{argmax}$ is not differentiable.


The structure of the BDC module, which determines the channel selection, is almost identical
to that of the BDP module except that the outputs of the softmax layer directly determine the 
channel selection $a^{k}_{i}$. First, the reshaped channel gain is fed into the BDC module 
which has $K\cdot N$ outputs. Then, the outputs of the BDC module are
divided into $N$ groups which are fed into $N$ independent softmax layer blocks
whose outputs are $a^{k}_{i}$. To be more specific, the $k$-th output
of the $i$-th softmax layer block can be interpreted as the probability that 
D2D TP $i$ selects channel $k$. Similar to the determination of the transmit power level, 
the determination of the channel selection is different for training and inference. Specifically, during training, 
the output of the BDC module, $a^{k}_{i}$, is directly used as the channel selection while during inference, 
we set $a^{k^*}_{i} = 1$ for $k^* = \underset{k}{\operatorname{argmax}} ~a^{k}_{i}$ and 
$a^{k}_{i} = 0$ for all other $k$ such that the constraint $a^{k}_{i} \in \{0, 1\}$ 
is satisfied in the actual implementation. Hence, due to this 
binarization, the resource allocation strategy employed during training may differ 
from that used for inference, which leads to performance deterioration. Thus, 
during the training process, the deviation of  $\hat{p}^{j}_{i}$ and $a^{k}_{i}$ 
from binary values is penalized, see Section IV, such that the binarization
error is reduced to zero.

\vspace{-2mm}
\subsection{Structure of Distributed DNN Model}\vspace{-1mm}

\begin{figure*}[t]
	\centerline{\includegraphics[width=16.0cm]{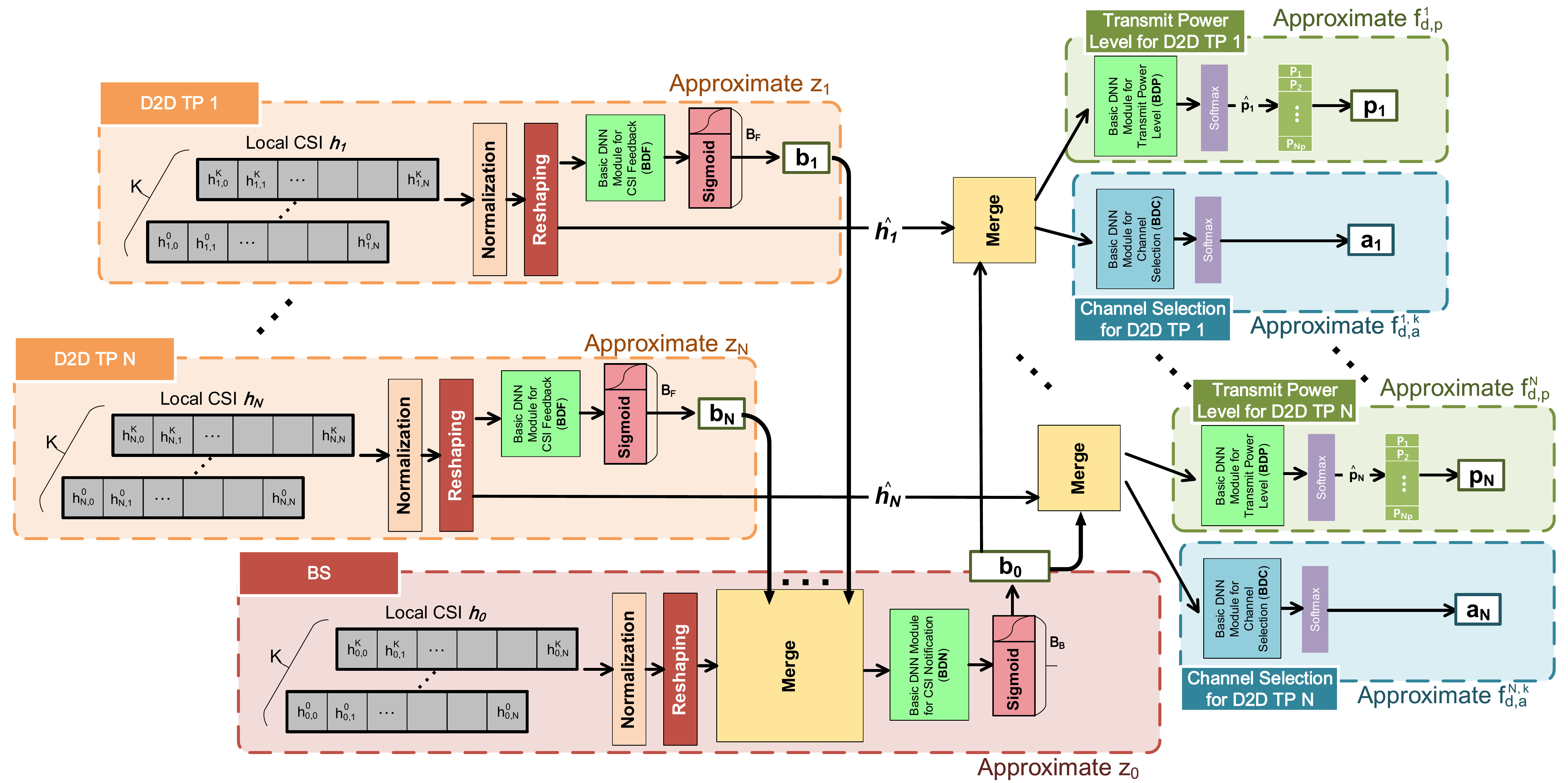}}\vspace{-2mm}
	\caption{Structure of distributed DNN model.}
	\label{fig_dist_dnn_structure}\vspace{-8mm}
\end{figure*}

In this subsection, we provide a detailed description of the structure of the
distributed DNN model which relies only on partial CSI. 
The complete structure of the distributed DNN model 
is depicted in Fig. \ref{fig_dist_dnn_structure}. Unlike the centralized 
case, where a single DNN module determines the resource allocation of all D2D TPs, 
in the distributed case, each D2D TP uses a different DNN module with the corresponding 
local CSI, $\bm{h}_{i}$, and summarized CSI, $\bm{b}_{0}$, as its input. 
Accordingly, the DNN module which determines $\bm{b}_{i}$, $\bm{a}_{i}$, 
and $p_{i}$ is implemented at the D2D TP $i$, $\forall i \in \mathbb{I}$,   
while the DNN module that determines $\bm{b}_{0}$
is implemented at the BS.

First, the local CSI is converted to the dB scale and normalized to have zero mean 
and unit variance, similar as for the centralized DNN structure.  
Let vector $\bm{\hat{h}}_{i}$ contain the pre-processed channel gains of D2D TP $i$. 
Then, $\bm{\hat{h}}_{i}$ is fed into the basic DNN module for CSI feedback (BDF), 
which has $B_{F}$ outputs and determines the encoded CSI 
that is sent to the BS, i.e., $\bm{b}_{i}$. The BDF module is followed by a 
sigmoid layer which converts input $x_{\textrm{sm}}$ to output  
$\frac{1}{1+e^{-x_{\textrm{sm}}}}$. The BDF module and sigmoid layer
together approximate the function for the encoding of the local CSI, 
$z_{i}(\cdot)$. Given that all outputs of the BDF module are subject to
separate sigmoid operations, the total number of outputs is
$B_{F}$, which is identical to the number of bits allocated for local 
CSI sharing. Unlike for the transmit power level and channel selection, for $\bm{b}_{i}$, multiple output values can have a value of 1, which is the reason for employing a sigmoid layer instead of a softmax layer.

After D2D TP $i$ has determined $\bm{b}_{i}$, this compressed CSI 
is sent to the BS and is merged to determine the encoded CSI which is broadcast 
back to all D2D TPs. To this end, the encoded CSI received 
from the D2D TPs and the local 
BS CSI are fed together into the basic DNN module for CSI notification (BDN) 
which determines $\bm{b}_{0}$, i.e., the BDN module approximates $z_{0}$. 
The output of the BDN module is fed into a sigmoid layer to obtain $\bm{b}_{0}$ 
which is then sent back to each D2D TP. Note that the output of all sigmoid layers, 
i.e., BDF and BDN, is forced to become either 0 or 1 during training to ensure proper binarization
as explained in Section IV.

Upon receiving $\bm{b}_{0}$ from the BS, each D2D TP
independently determines its transmit power level and channel selection
based on its local CSI, $\bm{h}_{i}$, and the summarized global CSI received from the BS, 
$\bm{b}_{0}$, using the BDP and BDC modules, as for the centralized DNN model. 
However, unlike for the centralized DNN model, where $N$ softmax layer blocks are used, 
only one softmax layer block is used for each D2D TP because the transmit
power level and the channel selection for only one D2D TP are determined.

\vspace{-1mm}
\section{Training and Inference of the Proposed DNN Models}\vspace{-1mm}

In this section, we first explain the training methodology of 
the proposed DNN models and the corresponding loss functions.
Then, we discuss how trained DNNs can be used 
to determine the resource allocation, i.e., the inference
of the DNN.

\vspace{-1mm}
\subsection{Training of the Proposed DNN Models}\vspace{-1mm}

In the proposed scheme, the constructed DNN models must be trained first, 
before they are used to allocate the wireless resources for the D2D TPs. 
We employ off-line training\footnote{In this paper, off-line
training is employed because training the DNN models is time consuming
and the speed of training can be improved through the use of the parallel computing. 
However, both the BS and the D2D TPs are unlikely to have this capability.}, i.e., the 
DNN models are trained on an independent computation unit and the parameters
of trained DNN models are forwarded to the BS and D2D TPs.
Herein, we adopt a hybrid training strategy exploiting both supervised 
and unsupervised learning \cite{Lee2018}. Specifically, the DNN models 
are first trained using a supervised learning strategy based on a few 
channel samples and the corresponding optimal resource allocation 
strategies.

This is referred to as coarse tuning (CT),
i.e., the DNN models are trained to mimic the optimal 
resource allocation policies which are given as a label data. 
After CT, a unsupervised learning strategy 
based on a large number of channel samples, which we refer to 
as fine tuning (FT), is applied to the DNNs which
have been already trained via CT. During FT, the 
weighted sum of objective functions
and the constraints in (\ref{prob_form_1}) and (\ref{prob_form_1_2})
are used as loss function. 
During both CT and FT, stochastic gradient descent algorithms, 
e.g., Adam (Adaptive Moment Estimation) \cite{Lee2018},
are used to update the weights of the DNNs.

The main rationale behind using the proposed hybrid training strategy is to combine the benefits of both individual training strategies, the optimality of supervised learning and the avoidance
of label data for unsupervised learning. More specifically, in unsupervised learning, the optimal resource allocation policy has to be found through trial-and-error and the algorithm may converge to a wrong policy. However, in supervised learning, the DNN is simply trained to mimic the provided optimal resource allocation, such that the optimal resource allocation policy can always be obtained. 
However, the acquisition of the labels needed
for supervised learning is time consuming 
because the label data, i.e., the optimal resource 
allocation, has to be found through an exhaustive search, 
while unsupervised learning only requires CSI for training.

In order to balance the trade-off between these two training strategies, in our proposed scheme, the DNN is first trained using CT, and then FT is used to finely tune the weights of the DNN models. Indeed CT can be considered as the initialization before the actual training starts, such that the weights of the DNN are adjusted to favorable initial values before FT is performed. As a result, the training time of the DNN for FT  is reduced compared to the case without CT, as can be observed from the simulation results provided in Section V.  Note that during CT, a small dataset is utilized in order to reduce the overhead introduced by computing the optimal resource allocation policy, which is needed as a label data during CT.

\vspace{-1mm}
\subsection{Loss Functions for Training}\vspace{-1mm}

During the CT phase, the DNN is trained to mimic the 
optimal channel and transmit power level selection, which 
we denote as $\acute{a}^{k}_{i}$ and $\acute{p}^{k}_{i}$, 
respectively, where $\acute{p}^{k}_{i} = 1$ only when $\textrm{P}_{k}$ 
is selected as desired power level. Note that $\acute{a}^{k}_{i}$ and $\acute{p}^{k}_{i}$
can be obtained through exhaustive search.
Then, the DNN models are trained to minimize 
loss functions, $\mathcal{L}^{\textrm{C}}_{\textrm{CT}}$ 
and $\mathcal{L}^{\textrm{D}}_{\textrm{CT}}$, for the centralized and distributed 
DNN models, respectively. The loss functions are given by
\begin{equation}
\begin{array}{lll}
\mathcal{L}^{\textrm{C}}_{\textrm{CT}} &=& -\sum^{N}_{i=1} \left( \sum^{K}_{k=1} \acute{a}^{k}_{i} \log(a^{k}_{i})
+\sum^{\textrm{N}_{\textrm{P}}}_{k=1} \acute{p}^{k}_{i} \log(\hat{p}^{k}_{i}) \right) + \rho^{\textrm{C}}_1 g^{\textrm{C}}_{1},\\
\mathcal{L}^{\textrm{D}}_{\textrm{CT}} &=& -\sum^{N}_{i=1} \left( \sum^{K}_{k=1} \acute{a}^{k}_{i} \log(a^{k}_{i})
+\sum^{\textrm{N}_{\textrm{P}}}_{k=1} \acute{p}^{k}_{i} \log(\hat{p}^{k}_{i}) \right) + \rho^{\textrm{D}}_1 g^{\textrm{D}}_{1} + \rho^{\textrm{D}}_2 g^{\textrm{D}}_{2},
  \label{eq_init_1}
\end{array}
\end{equation}
where $\rho^{\textrm{C}}_1$, $\rho^{\textrm{D}}_{1} $, and $\rho^{\textrm{D}}_{2}$ are control parameters which have to be positive. Furthermore, $g^{\textrm{C}}_{1}$, $g^{\textrm{D}}_{1}$, and $g^{\textrm{D}}_{2}$ in (\ref{eq_init_1}) are defined as follows:
\begin{equation}
\begin{array}{lll}
g^{\textrm{C}}_{1} &=&  -\sum \limits_{i \in \mathbb{I} }   \left( \sum \limits_{j =1}^{\textrm{N}_{\textrm{P}}}|\hat{p}^{j}_i - 0.5|^{\kappa} + \sum \limits_{k \in \mathbb{K} } |a_{i}^{k} - 0.5|^{\kappa} \right), \\
g^{\textrm{D}}_{1} &=&  -\sum \limits_{i \in \mathbb{I} }   \left( \sum \limits_{j =1}^{\textrm{N}_{\textrm{P}}}|\hat{p}^{j}_i - 0.5|^{\kappa} + \sum \limits_{k \in \mathbb{K} } |a_{i}^{k} - 0.5|^{\kappa} \right), \\
g^{\textrm{D}}_{2} &=&   -\sum \limits_{j=1}^{B_{F}}  \left( \sum \limits_{i \in \mathbb{I} }|b^{j}_{i} - 0.5|^{\kappa} + |b^{j}_{0} - 0.5|^{\kappa} \right),
  \label{eq_init_1_1}
\end{array}
\end{equation}
where $\kappa$ is a control parameter which has to be positive.

In (\ref{eq_init_1}), $-\sum^{N}_{i=1} \left( \sum^{K}_{k=1} \acute{a}^{k}_{i} \log(a^{k}_{i})
+\sum^{\textrm{N}_{\textrm{P}}}_{k=1} \acute{p}^{k}_{i} \log(\hat{p}^{k}_{i}) \right)$ 
corresponds to the cross entropy loss such that the value of the loss function is minimized when the outputs of the DNN, i.e., $a^{k}_{i}$ and $\hat{p}^{k}_{i}$, are identical to
the given label data, $\acute{a}^{k}_{i}$ and $\acute{p}^{k}_{i}$. Subsequently, 
the DNN model is trained to make $a^{k}_{i}$ and $\hat{p}^{k}_{i}$ identical to
the given optimal resource allocation, i.e., $\acute{a}^{k}_{i}$ and $\acute{p}^{k}_{i}$. We note that the use of the cross entropy loss function can improve the training speed \cite{Goodfellow2016} compared to the mean squared error loss function used in \cite{Sun2018}. Moreover, $g^{\textrm{C}}_{1}$ and $g^{\textrm{D}}_{1}$ are the binarization errors of the transmit power level and channel selection, i.e., $a^{k}_{i}$ and $\hat{p}^{k}_{i}$. Similarly, $g^{\textrm{D}}_{2}$ is the binarization error for the CSI encoding, i.e., $b^{j}_{i}$. Given that the values of $g^{\textrm{C}}_{1}$, $g^{\textrm{D}}_{1}$, and $g^{\textrm{D}}_{2}$  decrease as the $\hat{p}^{j}_i$, $a_{i}^{k}$, and $b^{j}_{i}$ approach either 0 or 1, the training pushes the values of $\hat{p}^{j}_i$, $a_{i}^{k}$, and $b^{j}_{i}$ towards binary values\footnote{For the binarization techniques considered in \cite{Lee2019c, Kim2018c}, a sigmoid function with steep slope is employed. This can possibly affect the performance of back-propagation based training because the gradient of a sigmoid function with steep slope is likely to be 0 for most input values. In our proposed scheme, by adopting a polynomial function in the loss function, the problem of vanishing gradient is avoided since the gradient of polynomial function is likely to be non-zero for all possible $a^{k}_{i}$ and $\hat{p}^{k}_{i}$.}. 
We note that the BDF, BDN, BDP, and BDC modules of the distributed DNN model
are jointly trained because the output of one module affects the other modules.


Next, we present the loss functions of the 
centralized and the distributed DNN models during the FT phase, $\mathcal{L}^{\textrm{C}}_{\textrm{FT}}$ 
and $\mathcal{L}^{\textrm{D}}_{\textrm{FT}}$, which are given as follows:
\begin{equation}
\begin{array}{lll}
\mathcal{L}^{\textrm{C}}_{\textrm{FT}} &=& - \hspace{-3mm}\sum \limits_{i \in \mathbb{I},  k \in \mathbb{K} } \textrm{SE}^{k}_{i} + \lambda^{\textrm{C}}_{1}  \sum \limits_{k \in \mathbb{K} } \dfrac{ \max( \textrm{SE}_{\textrm{thr}} - \textrm{SE}^{k}_{0}, 0)}{\textrm{SE}_{\textrm{thr}} +\delta_{FT}} + \lambda^{\textrm{C}}_{2} g^{\textrm{C}}_{1},\\
\mathcal{L}^{\textrm{D}}_{\textrm{FT}} &=& - \hspace{-3mm} \sum \limits_{i \in \mathbb{I},  k \in \mathbb{K} } \tilde{\textrm{SE}}^{k}_{i} + \lambda^{\textrm{D}}_{1}  \sum \limits_{k \in \mathbb{K} } \dfrac{ \max(  \textrm{SE}_{\textrm{thr}} - \tilde{\textrm{SE}}^{k}_{0}  , 0)}{\textrm{SE}_{\textrm{thr}}+\delta_{FT}} + \lambda^{\textrm{D}}_{2} g^{\textrm{D}}_{1} + \lambda^{\textrm{D}}_{3} g^{\textrm{D}}_{2},
  \label{eq_loss_SE_dist}
\end{array}
\end{equation}
where $\lambda^{\textrm{C}}_{1}$, $\lambda^{\textrm{C}}_{2}$, $\lambda^{\textrm{D}}_{1}$, 
$\lambda^{\textrm{D}}_{2}$, and $\lambda^{\textrm{D}}_{3}$ are positive control parameters
and $\delta_{FT}$ is a small positive constant which
prevents division by zero when $\textrm{SE}_{\textrm{thr}} = 0$.

In (\ref{eq_loss_SE_dist}), the first term corresponds to the negative objective function 
used for resource allocation, namely, sum SE of the D2D TPs. The value of 
the loss function decreases when the sum SE of the D2D TPs increases. 
Moreover, the second term in (\ref{eq_loss_SE_dist}) depends on how much 
the QoS constraint for legacy cellular uplink transmission is violated, 
i.e., $\textrm{SE}^{k}_{0}, \tilde{\textrm{SE}}^{k}_{0}  < \textrm{SE}_{\textrm{thr}}$. Accordingly, the 
value of the loss function increases when the QoS constraint is not satisfied.
As a result, the DNNs can be trained when the optimal 
resource allocation is not available as label data. Moreover, as in 
(\ref{eq_init_1}), $g^{\textrm{C}}_{1}$, $g^{\textrm{D}}_{1}$, and $g^{\textrm{D}}_{2}$ 
are used for binarization. Note that the values of 
$\lambda^{\textrm{C}}_{1}$, $\lambda^{\textrm{C}}_{2}$, $\lambda^{\textrm{D}}_{1}$, 
$\lambda^{\textrm{D}}_{2}$, and $\lambda^{\textrm{D}}_{3}$ determine whether
the DNN model puts more emphasis on the maximization of the SE, meeting the QoS constraint,
or the binarization of the output of the DNN.

\hspace{-0.25cm}\textbf{Remark 1}
Although the proposed loss functions in (\ref{eq_loss_SE_dist}) are different from the conventional loss functions 
typically used for DL, e.g., the cross entropy loss, the back propagation 
based training of the DNN is still applicable because the proposed loss functions are 
differentiable \cite{Gu2015}.

\hspace{-0.25cm}\textbf{Remark 2}
The proposed DNN model can cope with different design objectives and constraints easily by  
modifying the loss function. For example, the DNN model can be trained
to maximize the EE by replacing $\textrm{SE}^{k}_{i}$ in (\ref{eq_loss_SE_dist}) by $\textrm{EE}^{k}_{i}$, which is given by
$\textrm{EE}^{k}_{i} =  \frac{\textrm{SE}^{k}_{i}}{p_{i} + \textrm{P}_{\textrm{CIR}}}$,
where $\textrm{P}_{\textrm{CIR}}$ is the circuit power consumption of the D2D users.

\hspace{-0.25cm}\textbf{Remark 3}
In the proposed scheme, the training is performed before the DNNs are 
employed in the communication system, i.e., in an off-line manner. Hence, the  
overhead incurred during training of the DNNs, which can be huge, 
is not an obstacle for real-time operation of the proposed scheme 
\cite{Lee2018, Lee2018b, Lee2019a, Lee2020}.

\vspace{-1mm}
\subsection{Inference of the Proposed DNN Model}\vspace{-1mm}

The transmit power level ($p_{i}$), channel selection ($a_{i}^{k}$), 
and CSI feedback ($\bm{b}_{i}$, $\bm{b}_{0}$) can be determined 
by feeding the current channel gains to the trained DNN models. 
Given that the proposed DNN models involve only simple matrix calculations and
functions, the optimal resource allocation policy
can be determined fast even when the number 
of users and channels is large, as will be confirmed later in Section V.

\vspace{-1mm}
\section{Performance Evaluation} \vspace{-1mm}

\begin{figure}
\centerline{\includegraphics[width=7.5cm]{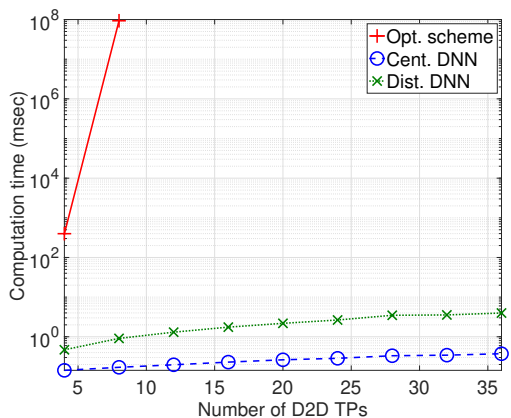}}\vspace{-1mm}
 \caption{Computation time of optimal and proposed scheme vs. $N$.}
\label{fig_comp_time}\vspace{-5mm}
\end{figure}

In this section, we assess the performance of the proposed DL-based
resource allocation scheme. We assume that the users are randomly 
distributed over a 100 m $\times$ 100 m area, where the maximum distance between a 
transmitter and a receiver in the same TP is set to 30 m. 
Moreover, we assume that $\textrm{P}_{\textbf{\textrm{M}}}$ = 23 dBm and 
$\textrm{N}_{\textrm{P}}$ = 8, i.e., the transmit power is equally divided into 8 levels.
Accordingly, $\textrm{P}_{0} = 0$, $\textrm{P}_{1} = 28.57$ mW
$\textrm{P}_{2} = 57.14$ mW, $\textrm{P}_{3} = 85.71$ mW, 
$\textrm{P}_{4} = 114.28$ mW,  $\textrm{P}_{5} = 142.85$ mW, 
$\textrm{P}_{6} = 171.42$ mW,  and $\textrm{P}_{7} = 200$ mW. 
Furthermore, $N = 3$, $K = 3$,\footnote{In the default simulation 
environment, we have considered small $N$ and $K$ due 
to the excessive computation time needed to determine the optimal performance 
via exhaustive search even for moderate values of $N$ and $K$, see Fig. \ref{fig_comp_time}.} 
$W =$ 10 MHz, $N_{0}$ = -173 dBm/Hz,
$B_{F}$ = 12, and $B_{B}$ = 24, such that the CSI is compressed to 12 bits and 24 bits
for the D2D TPs and the BS, respectively. A simplified path loss model with 
path loss coefficient $10^{3.453}$ and path loss exponent $3.8$ is 
adopted \cite{Lee2018}. The small-scale fading gains are modeled as 
independent and identically distributed (i.i.d.) circularly symmetric 
complex Gaussian (CSCG) random variables with zero mean 
and unit variance. Furthermore, 
we assume that $5 \times 10^{7}$ and $10^{5}$ channel samples are 
generated for training and evaluation, respectively. In addition, Adam \cite{Kingma2014} is 
used for training where the learning rate, which determines
the speed of learning,  is set to $10^{-3}$ and $3 \times 10^{-6}$ for the 
CT and FT phases, respectively.

Regarding the structure of the DNN model, for the centralized DNN model, we assume 
that the numbers of hidden layers and hidden nodes for both the BDP and BDC modules  are 16 
and 400, respectively. Moreover, for the distributed DNN model, the numbers of 
hidden layers and hidden nodes of the BDF, BDN, BDP, and BDC modules  are set to 8 
and 150, respectively. Furthermore, the values of the control parameters
for the QoS constraint and binarization, i.e., 
$\rho^{\textrm{C}}_1$, $\rho^{\textrm{D}}_{1} $,
$\rho^{\textrm{D}}_{2}$, $\lambda^{\textrm{C}}_{1}$, $\lambda^{\textrm{C}}_{2}$, $\lambda^{\textrm{D}}_{1}$, 
$\lambda^{\textrm{D}}_{2}$, and $\lambda^{\textrm{D}}_{3}$
are adaptively determined for each simulation environment
through trial-and-error. For performance evaluation, three conventional 
schemes are considered, namely 1) the optimal scheme, 2) the naive distributed DL-based scheme 
proposed in \cite{Lee2018}, and 3) a random scheme. For the optimal scheme, the optimal 
resource allocation policy is found through exhaustive search. For our default simulation 
setting, where $\textrm{N}_{\textrm{P}}$ = 8, $N = 3$, and $K = 3$, the total number of possible
resource allocation strategies is 10648, all of which have to be examined in 
the exhaustive search. For the naive scheme, the centralized DNN model, which is 
trained based on the complete CSI, is used for distributed operation, where the average values of the 
non-local CSI values are employed to determine the resource allocation.
For the random scheme, the channel and 
transmit power levels are randomly selected. 
\begin{figure}
\centerline{\includegraphics[width=7.5cm]{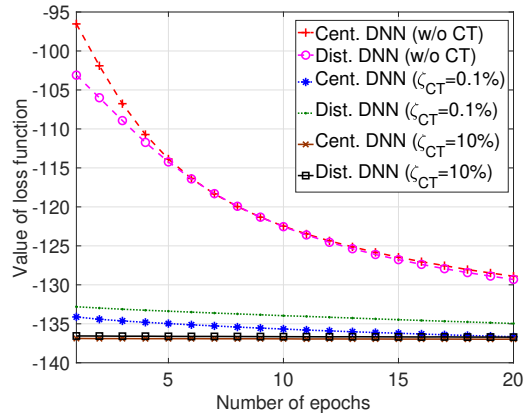}}\vspace{-2mm}
 \caption{Convergence of loss for different numbers of samples used in CT phase.}
\label{fig_convergence}\vspace{-5mm}
\end{figure}

First, in order to confirm the benefits of the proposed hybrid training strategy,
where CT is used as a means to initialize the DNN, we evaluate 
the evolution of the loss functions for FT, 
$\mathcal{L}^{\textrm{C}}_{\textrm{FT}}$ 
and $\mathcal{L}^{\textrm{D}}_{\textrm{FT}}$, 
by varying the proportion of samples used for CT, 
$\zeta_{\textrm{CT}}$, as depicted
in Fig. \ref{fig_convergence}, where an epoch is 
the number times that the entire training dataset
is used for training. As can be 
observed from Fig. \ref{fig_convergence}, the speed of convergence 
can be greatly improved by employing CT, even when only a 
small fraction of samples is used for CT, i.e., 
$\zeta_{\textrm{CT}} = 0.1 \%$, which demonstrates 
the usefulness of CT. Moreover, for the distributed DNN model, more time is required
for training than for the centralized DNN model because the 
encoding of the CSI has to be also optimized in addition 
to the resource allocation. We note that 
the training time required for CT is negligible
because the amount of training data is small.  For example, 
for $\zeta_{\textrm{CT}} = 0.1 \%$, CT takes 0.04 seconds
while FT takes more than 60 seconds for one epoch of training.
\begin{figure*}[t!]
	\begin{center}

	\subfigure[SE of D2D TPs when $\textrm{SE}_{\textrm{thr}}$ = 1 b/s/Hz.]{
		\includegraphics[width=7.5cm]{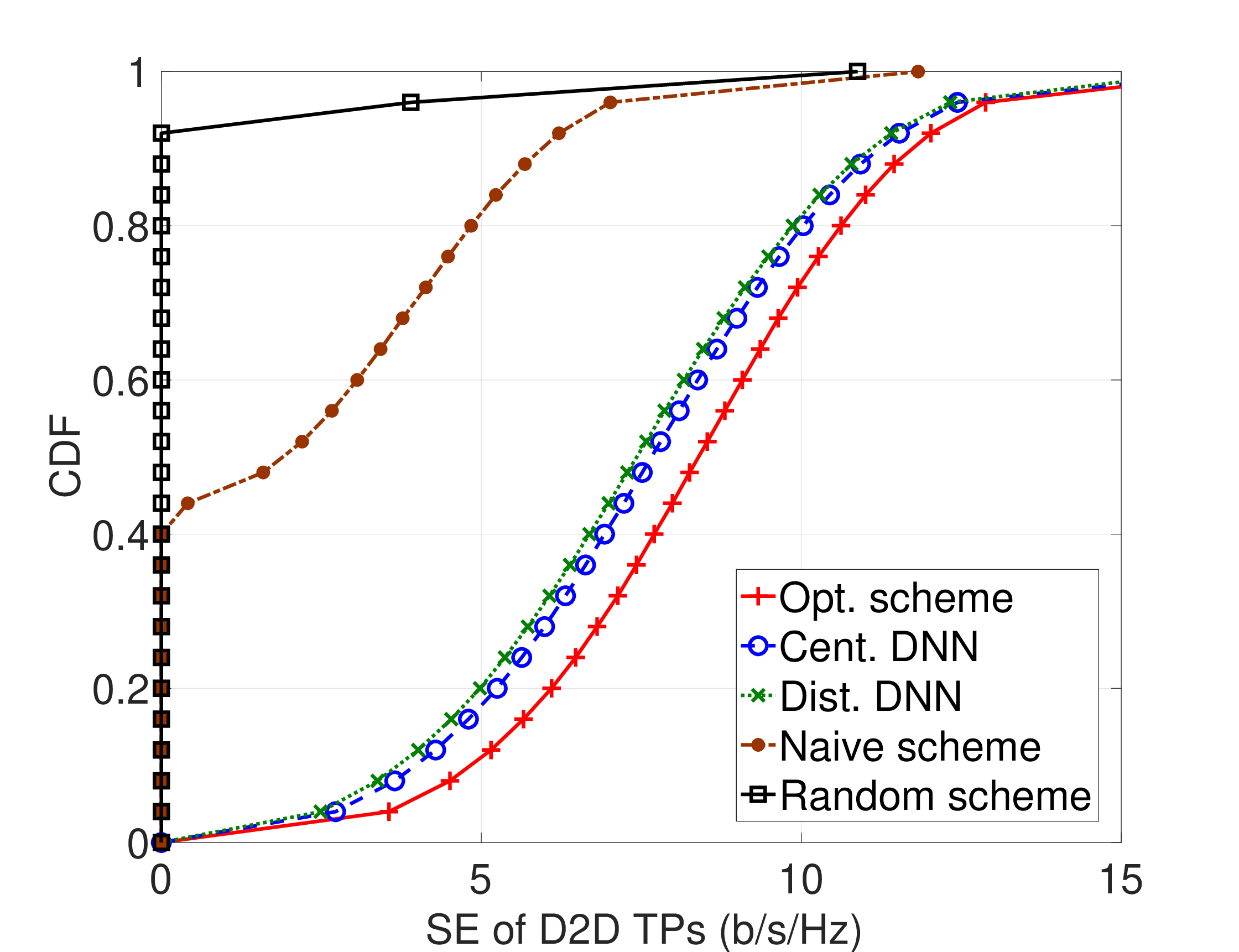}
		\label{rate_fig_a}
	}
	\subfigure[SE of D2D TPs when $\textrm{SE}_{\textrm{thr}}$ = 0 b/s/Hz.]{
		\includegraphics[width=7.5cm]{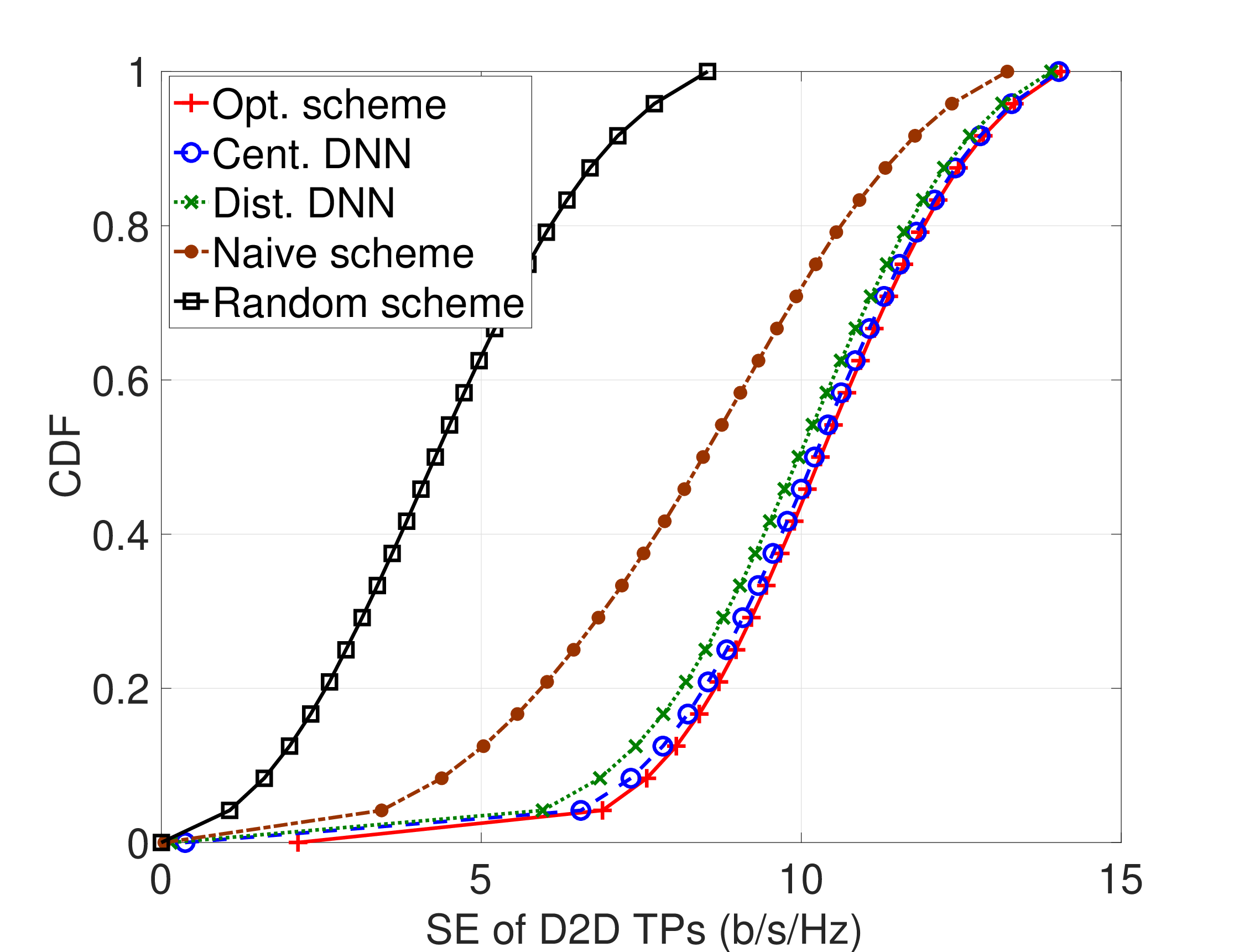}
		\label{rate_fig_b}
	}
	\end{center}\vspace{-4mm}
	\caption{CDF of SE of D2D TPs for different values of $\textrm{SE}_{\textrm{thr}}$.}\vspace{-5mm}
	\label{CDF_SE_fig}
\end{figure*}
\begin{figure*}[t!]
	\begin{center}

	\subfigure[SE of CUEs when $\textrm{SE}_{\textrm{thr}}$ = 1 b/s/Hz.]{
		\includegraphics[width=7.5cm]{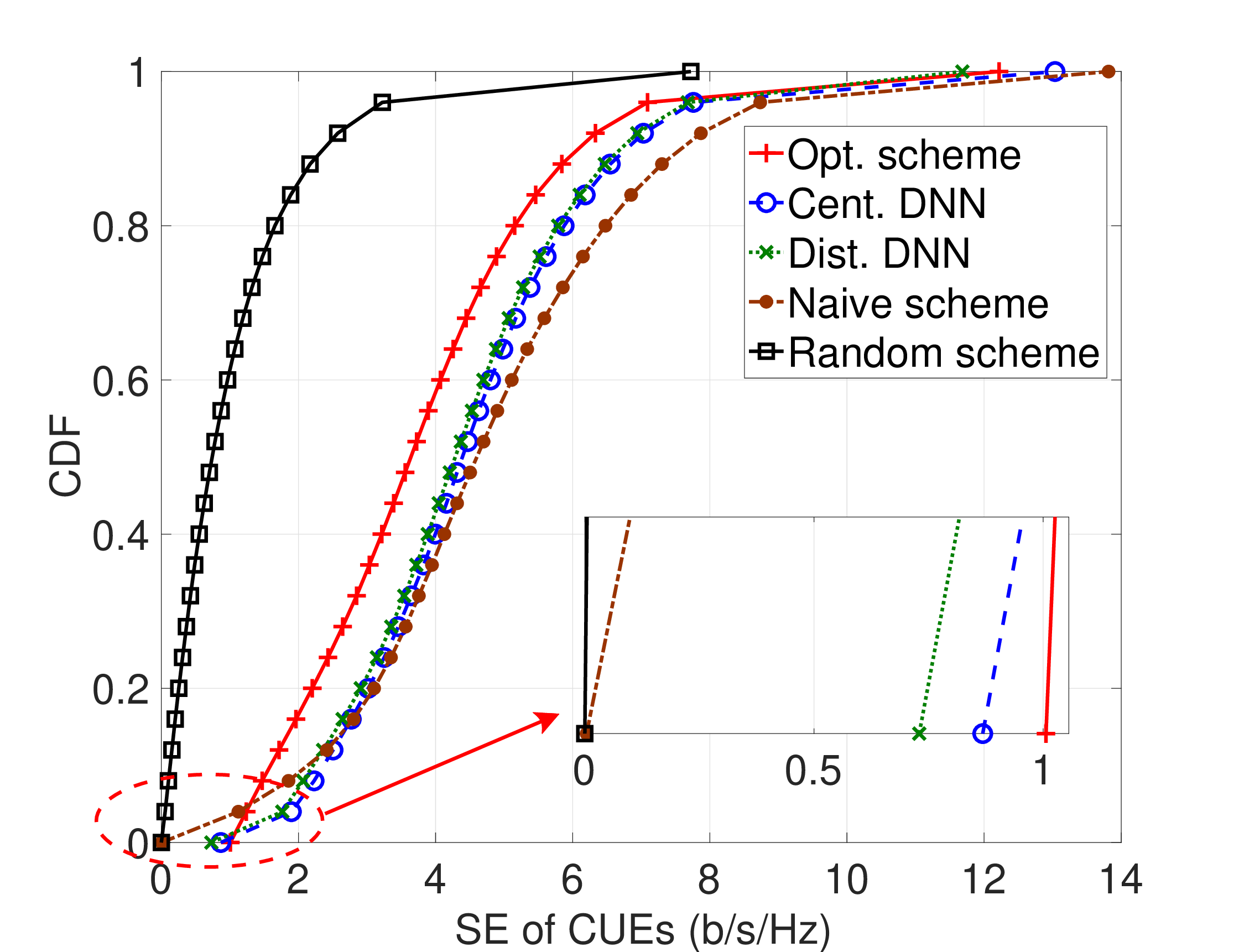}
		\label{cue_fig_a}
	}
	\subfigure[SE of CUEs when $\textrm{SE}_{\textrm{thr}}$ = 0 b/s/Hz.]{
		\includegraphics[width=7.5cm]{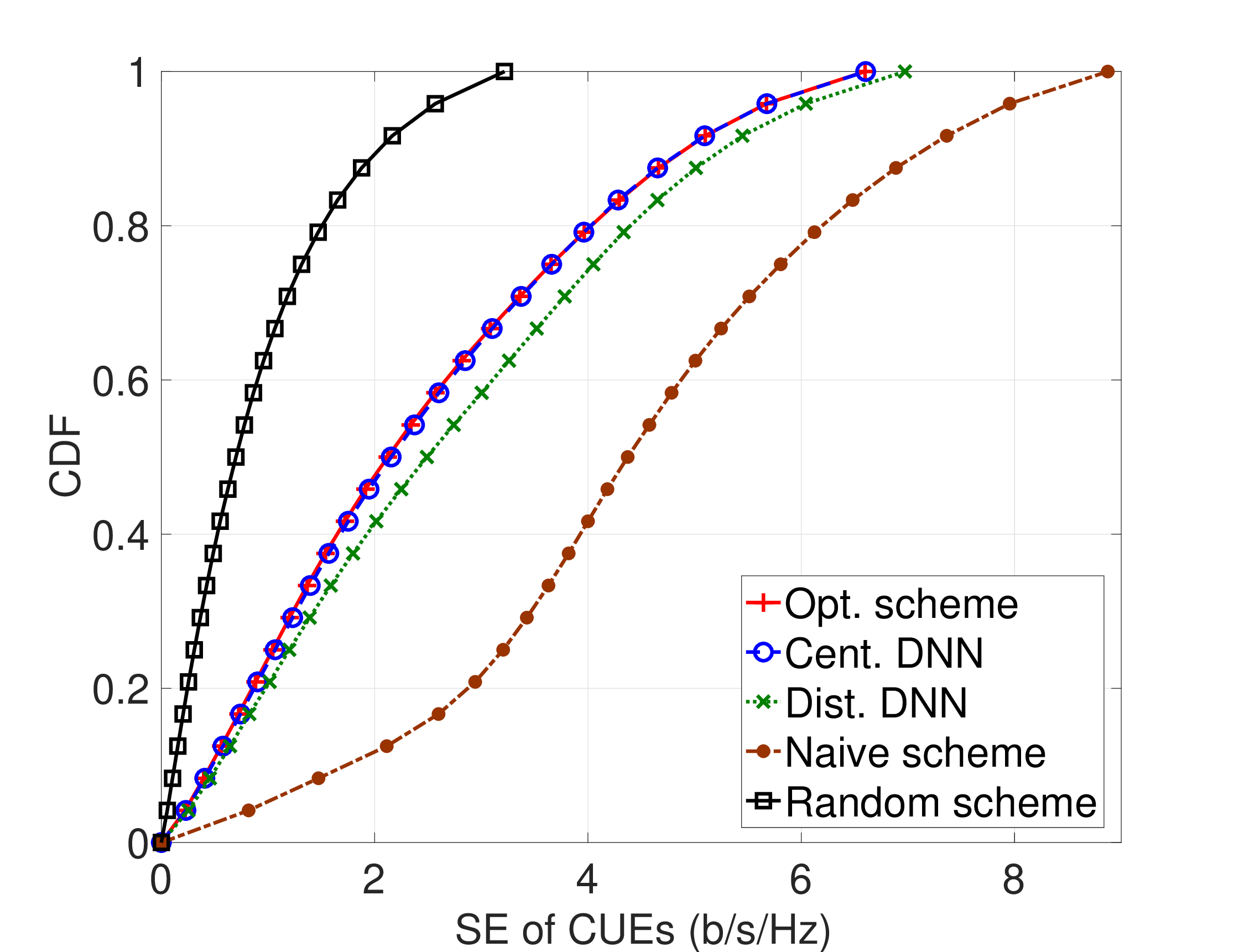}
		\label{cue_fig_b}
	}
	\end{center}\vspace{-4mm}
	\caption{CDF of SE of CUEs for different values of $\textrm{SE}_{\textrm{thr}}$.}\vspace{-5mm}
	\label{CDF_CUE_fig}
\end{figure*}


In Figs. \ref{CDF_SE_fig} and \ref{CDF_CUE_fig}, we show the cumulative distribution 
function (CDF) of the SE of the D2D TPs and the CUEs for different $\textrm{SE}_{\textrm{thr}}$.
For the calculation of the SE, we set the SE of the D2D TPs to zero when the QoS constraint 
is not satisfied in order to penalize the QoS violation.
From the simulation results, we observe that the SE of the D2D TPs deteriorates
while that of the CUEs improves when $\textrm{SE}_{\textrm{thr}}$ increases because 
the transmit power of the D2D TPs is more severely regulated when $\textrm{SE}_{\textrm{thr}}$
is larger in order to meet the QoS constraints of the CUEs. As can be observed from
the results in Fig. \ref{CDF_SE_fig}, the performance of DNN-based resource allocation 
is almost identical to that of the optimal scheme when $\textrm{SE}_{\textrm{thr}} = 0$. 
Although the SE of the D2D TPs for the proposed scheme is slightly degraded when 
$\textrm{SE}_{\textrm{thr}} =$ 1 b/s/Hz, it is much higher than that obtained for the
naive and random schemes. Furthermore, we observe that the SE of the D2D TPs is slightly 
lower for the distributed DNN model 
than for the centralized DNN model, because for the 
distributed scheme, only partial CSI is employed. In addition, the random 
scheme has the worst SE for the D2D TPs and the CUEs in all cases, which 
underscores the need for a proper resource allocation optimization.

\begin{figure*}[t!]
	\begin{center}

	\subfigure[CDF of binarization error for channel selection and transmit power level.]{
		\includegraphics[width=7.5cm]{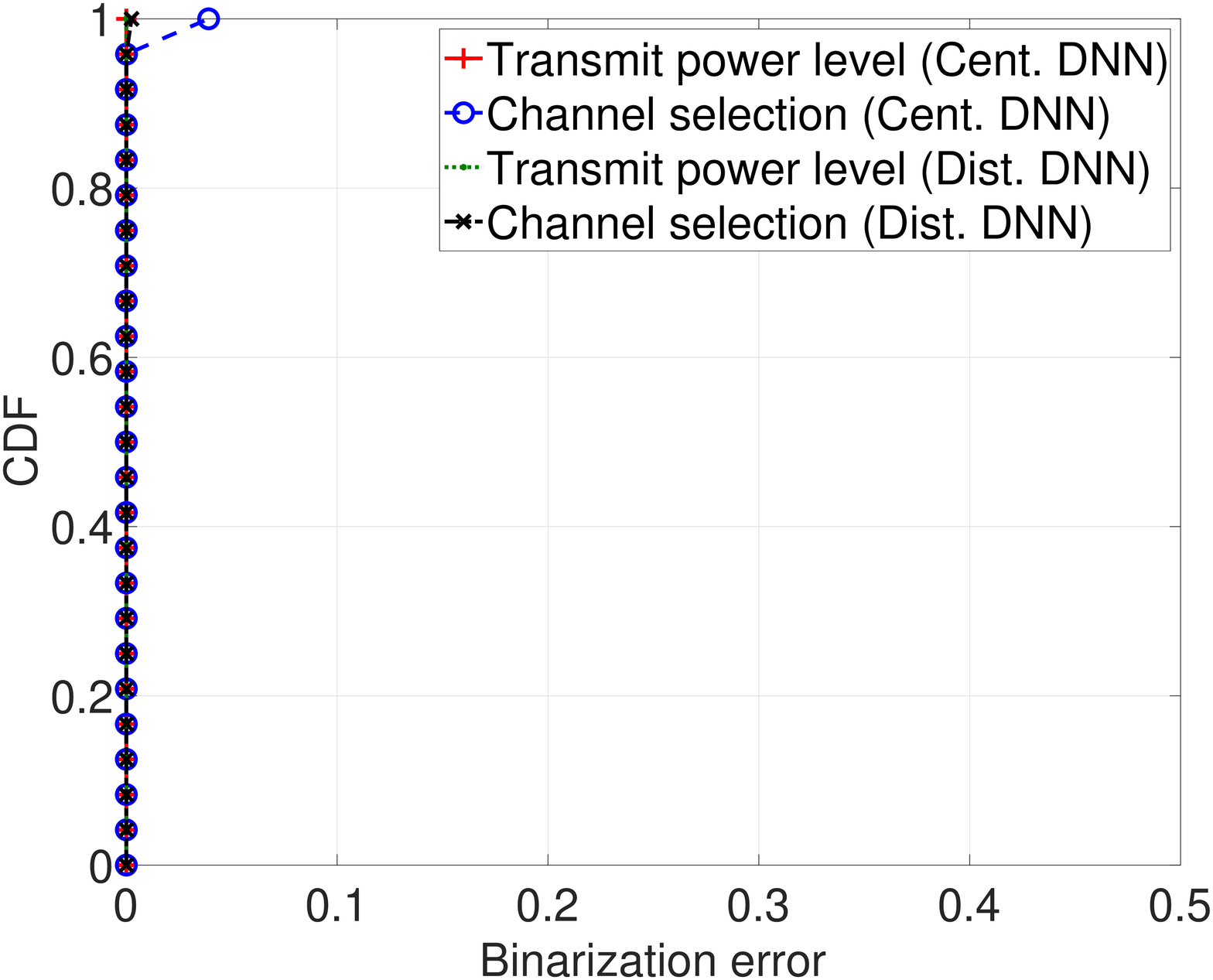}
		\label{fig_CDF_CS_PL}
	}
	\subfigure[CDF of binarization error for CSI feedback.]{
		\includegraphics[width=7.5cm]{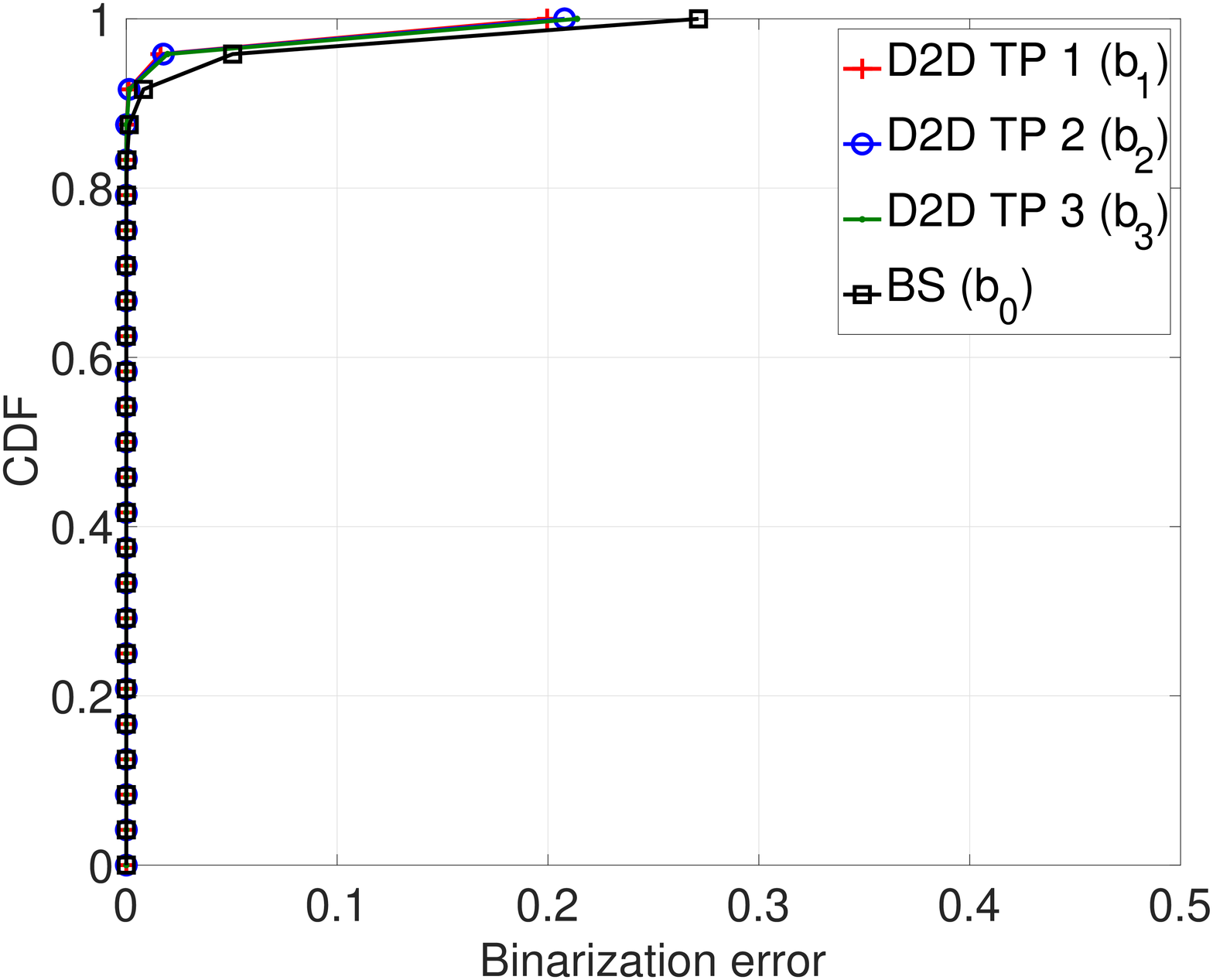}
		\label{fig_CDF_CSI_exchange}
	}
	\end{center}\vspace{-4mm}
	\caption{CDF of binarization error for outputs of DNN models.}\vspace{-5mm}
	\label{fig_CDF_CS_PL_CSI_exchange}
\end{figure*}

\begin{figure*}[t!]
	\begin{center}

	\subfigure[Average SE of D2D TPs.]{
		\includegraphics[width=5.1cm]{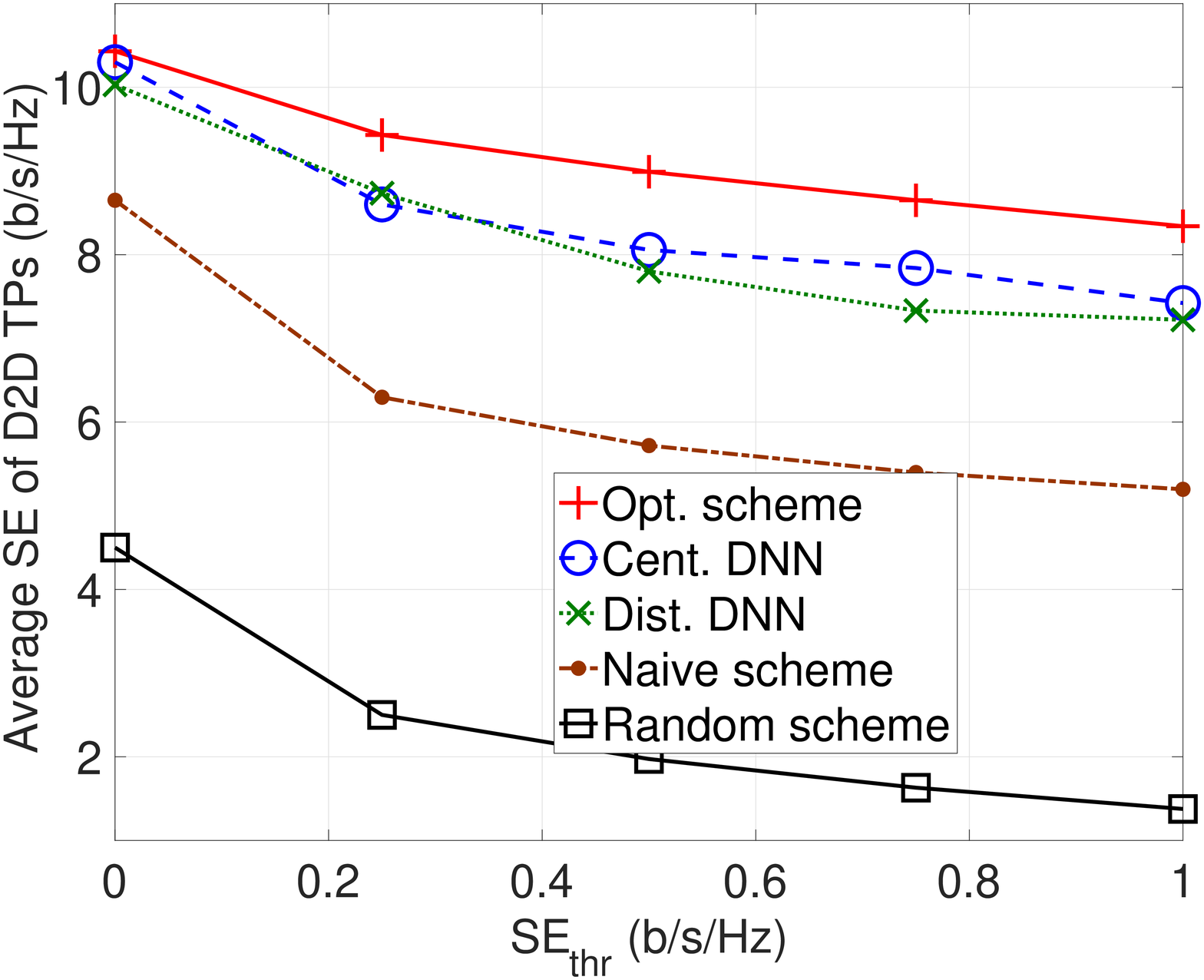}
		\label{SE_THR_fig_a}
	}
	\subfigure[Probability of QoS violation.]{
		\includegraphics[width=5.1cm]{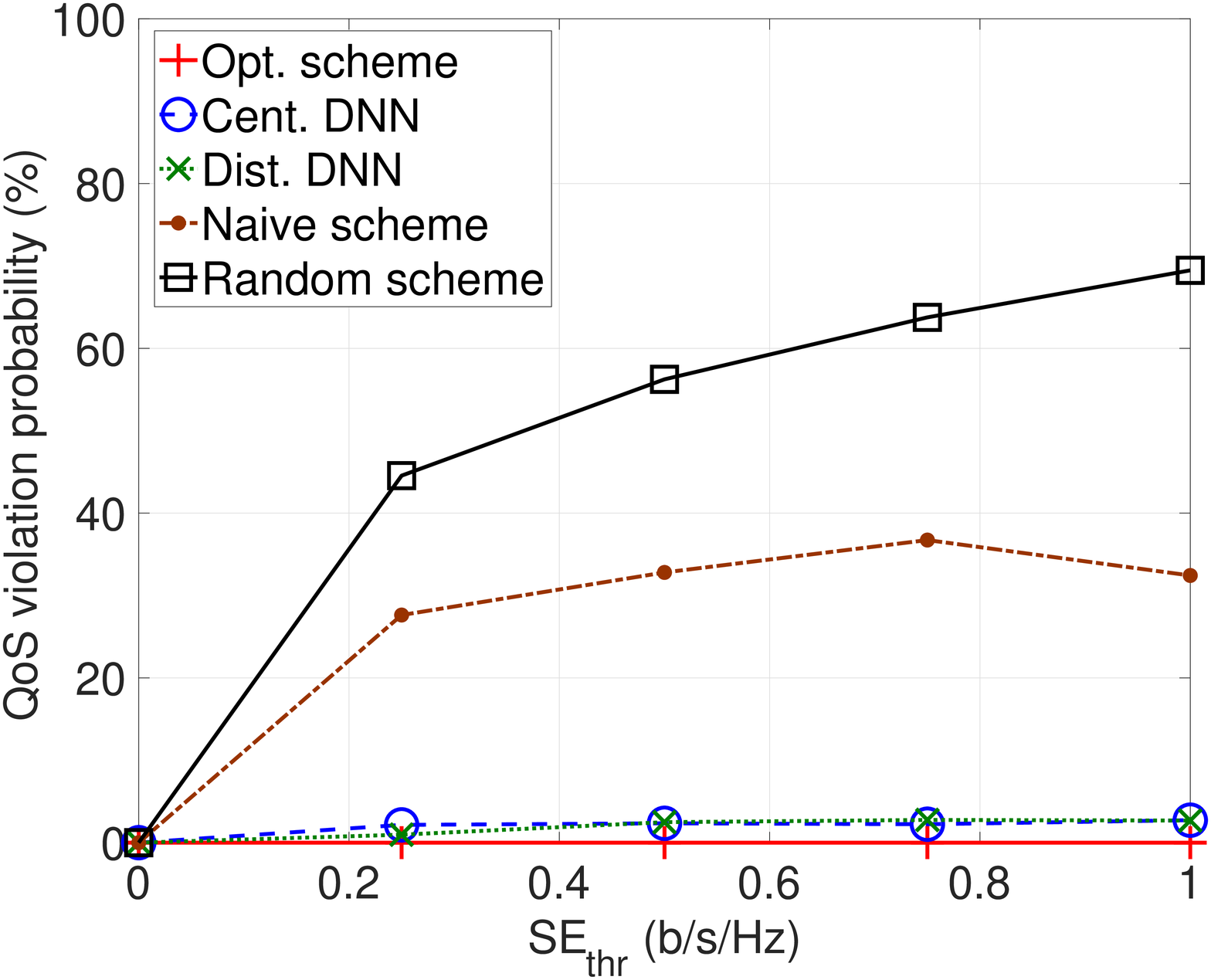}
		\label{SE_THR_fig_b}
	}
	\subfigure[Level of QoS violation.]{ 
		\includegraphics[width=5.1cm]{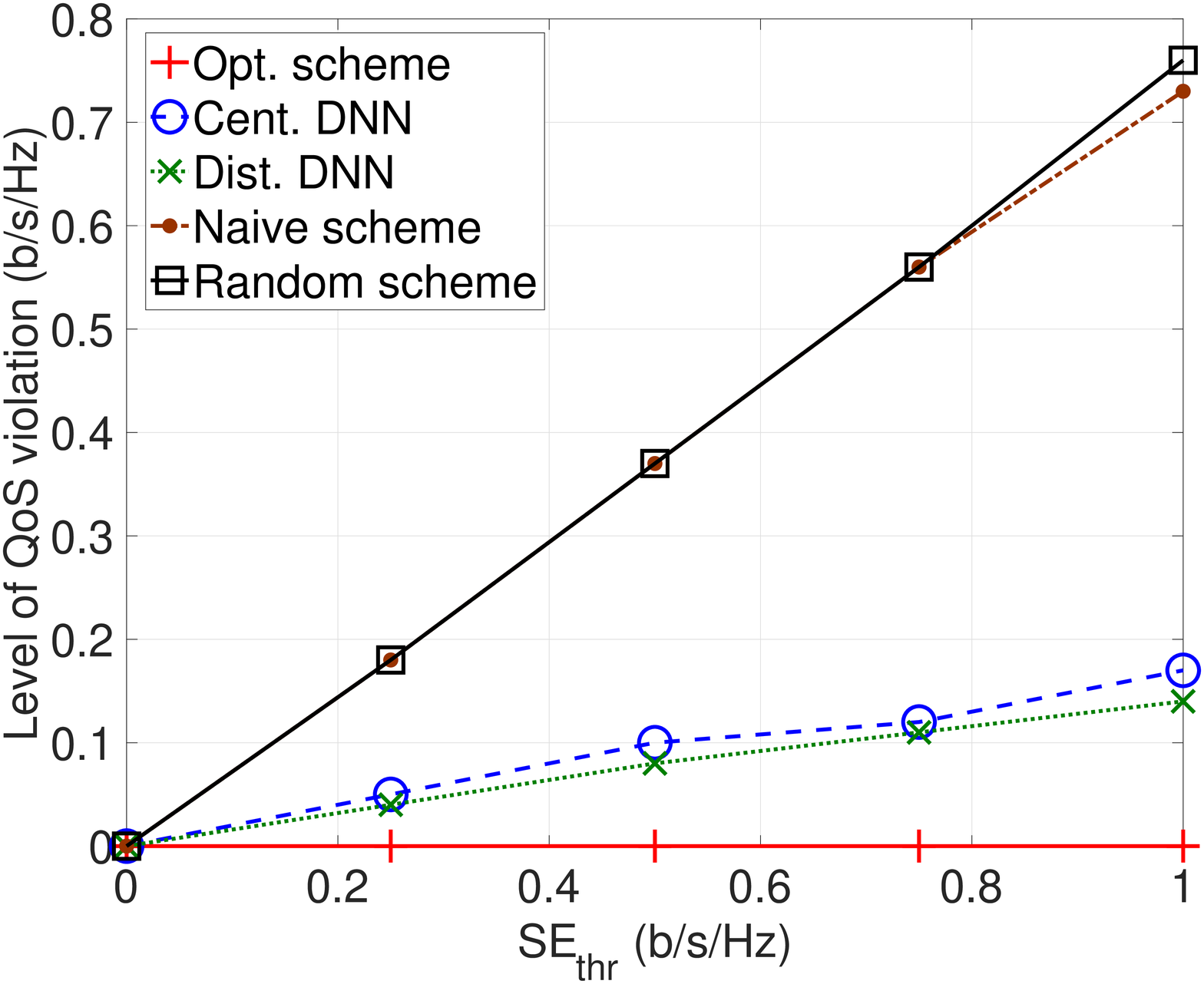}
		\label{SE_THR_fig_c}
	}
	\end{center}\vspace{-4mm}
	\caption{Performance of proposed schemes as a function of $\textrm{SE}_{\textrm{thr}}$.}\vspace{-5mm}
	\label{SE_THR_fig}
\end{figure*}

The naive scheme yields a poor SE for the D2D TPs, especially 
when $\textrm{SE}_{\textrm{thr}}$ = 1 b/s/Hz. This is mainly due 
to high probability that the QoS constraint is violated\footnote{The violation of the QoS constraint 
is obvious from the starting point of the CDF of the SE of the CUEs
in Fig. \ref{CDF_CUE_fig}. More specifically, the CDF of the optimal scheme 
starts at $\textrm{SE}_{\textrm{thr}}$ such that the SE of the CUEs
is always larger than $\textrm{SE}_{\textrm{thr}}$. On the other hand,
the CDF of the other schemes starts at a point which is smaller than 
$\textrm{SE}_{\textrm{thr}}$ such that some CUEs will achieve a
lower SE than $\textrm{SE}_{\textrm{thr}}$, i.e., the QoS constraint 
is violated. We note that the starting point 
of the CDF of the proposed scheme is much closer to that of the optimal scheme compared to the other schemes, 
i.e., the QoS constraint is almost always satisfied.} for the naive 
scheme. This probability can be as high as 32\% when $\textrm{SE}_{\textrm{thr}}$ = 1 b/s/Hz,
cf. Fig. \ref{SE_THR_fig_b}. We note that for the proposed centralized and distributed DNN-based schemes,
the QoS violation probability is less than 2.7\%, 
which validates the usefulness of the proposed design, see Fig. \ref{SE_THR_fig_b}.
Moreover, we observe that although the QoS constraint may be violated for some channel
realization for the proposed scheme, the overall SE of the CUEs is increased compared to the 
optimal scheme, cf. Fig. \ref{CDF_CUE_fig}. This behavior is due to the fact
that the proposed scheme restricts the transmit power of the D2D TPs more severely 
such that the overall SE of the D2D TPs is slightly degraded but that of the CUEs is slightly 
increased compared to the optimal scheme.

In order to confirm that the outputs of the DNNs are properly binarized, 
we consider the CDF of the binarization error defined as 
$|\lfloor x \rceil - x|$, where $\lfloor \cdot \rceil$ denotes
the nearest integer value for $\textrm{SE}_{\textrm{thr}}$ = 1 b/s/Hz.  
The CDF of the errors for the transmit power level
and channel selection are shown in Fig. \ref{fig_CDF_CS_PL} and those for
$\bm{b}_{1}$ (CSI from D2D TP 1), $\bm{b}_{2}$ (CSI from D2D TP 2), $\bm{b}_{3}$ (CSI from D2D TP 3), 
and $\bm{b}_{0}$ (CSI from BS) are depicted in Fig. \ref{fig_CDF_CSI_exchange}.
Given that these optimization variables are in the range between 0 and 1, the maximum value
of the binarization error is 0.5. From Fig. \ref{fig_CDF_CS_PL_CSI_exchange}, we observe that the 
binarization error for all optimization variables is negligibly small, which confirms that
our DNN models are able to generate binary values.


Next, we show the average SE of the D2D TPs, the probability of QoS constraint violation, 
and the level of QoS violation, which is defined as the difference between the QoS requirement
and the SE of the CUEs when the QoS constraint is violated, i.e., 
$\mathbb{E}_{\textrm{SE}^{k}_{0}}\left[\textrm{SE}_{\textrm{thr}} - \textrm{SE}^{k}_{0} ~|~ \textrm{SE}_{\textrm{thr}} > \textrm{SE}^{k}_{0}
 \right]$, as a function of  $\textrm{SE}_{\textrm{thr}}$ in 
Figs. \ref{SE_THR_fig_a} - \ref{SE_THR_fig_c}. As can be observed from 
Fig. \ref{SE_THR_fig_a}, the average SE of the D2D TPs decreases as $\textrm{SE}_{\textrm{thr}}$
increases because the transmit power and channel usage of the D2D TPs have to be 
restricted more severely in order to satisfy the QoS constraints of the CUEs. Moreover, we observe 
that the average SE of the proposed scheme is slightly degraded compared to 
the optimal scheme while the gap between both schemes decreases 
for small $\textrm{SE}_{\textrm{thr}}$. In Fig. \ref{SE_THR_fig_b}, we observe
that the QoS violation probability of the
proposed schemes is close to 0 while that of the conventional schemes
is much higher, which highlights the effectiveness of the proposed scheme.
Moreover, from Fig. \ref{SE_THR_fig_c}, we observe that
the level of QoS violation for the proposed scheme is small such that
even though the QoS constraint is violated, the SE of the CUEs is likely very 
close to $\textrm{SE}_{\textrm{thr}}$ such that the impact of the QoS
violation is minimal.
\begin{figure*}[t!]
	\begin{center}

	\subfigure[Average SE of D2D TPs.]{
		\includegraphics[width=5.1cm]{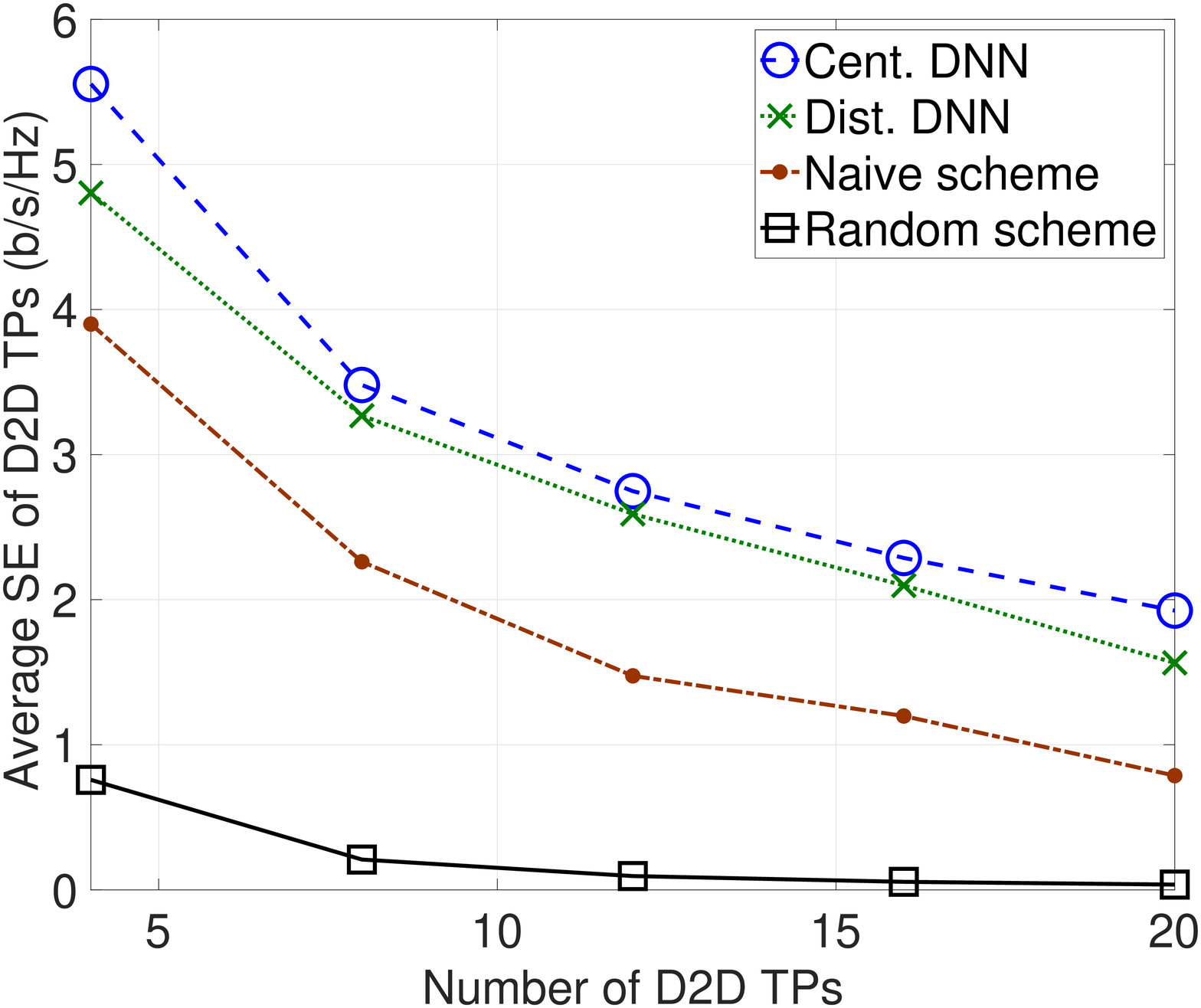}
		\label{SE_DUE_fig_a}
	}
	\subfigure[Probability of QoS violation.]{
		\includegraphics[width=5.1cm]{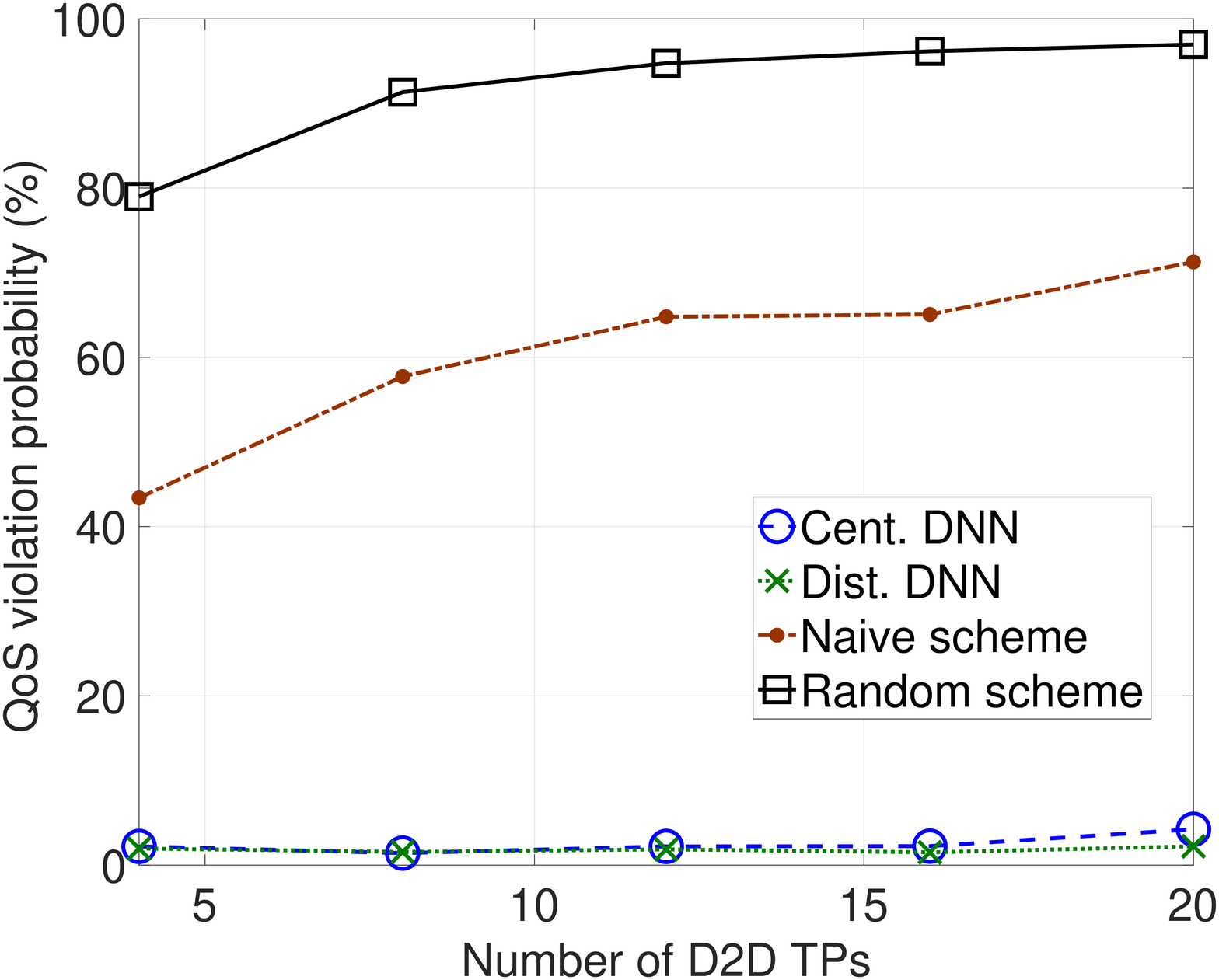}
		\label{SE_DUE_fig_b}
	}
	\subfigure[Level of QoS violation.]{ 
		\includegraphics[width=5.1cm]{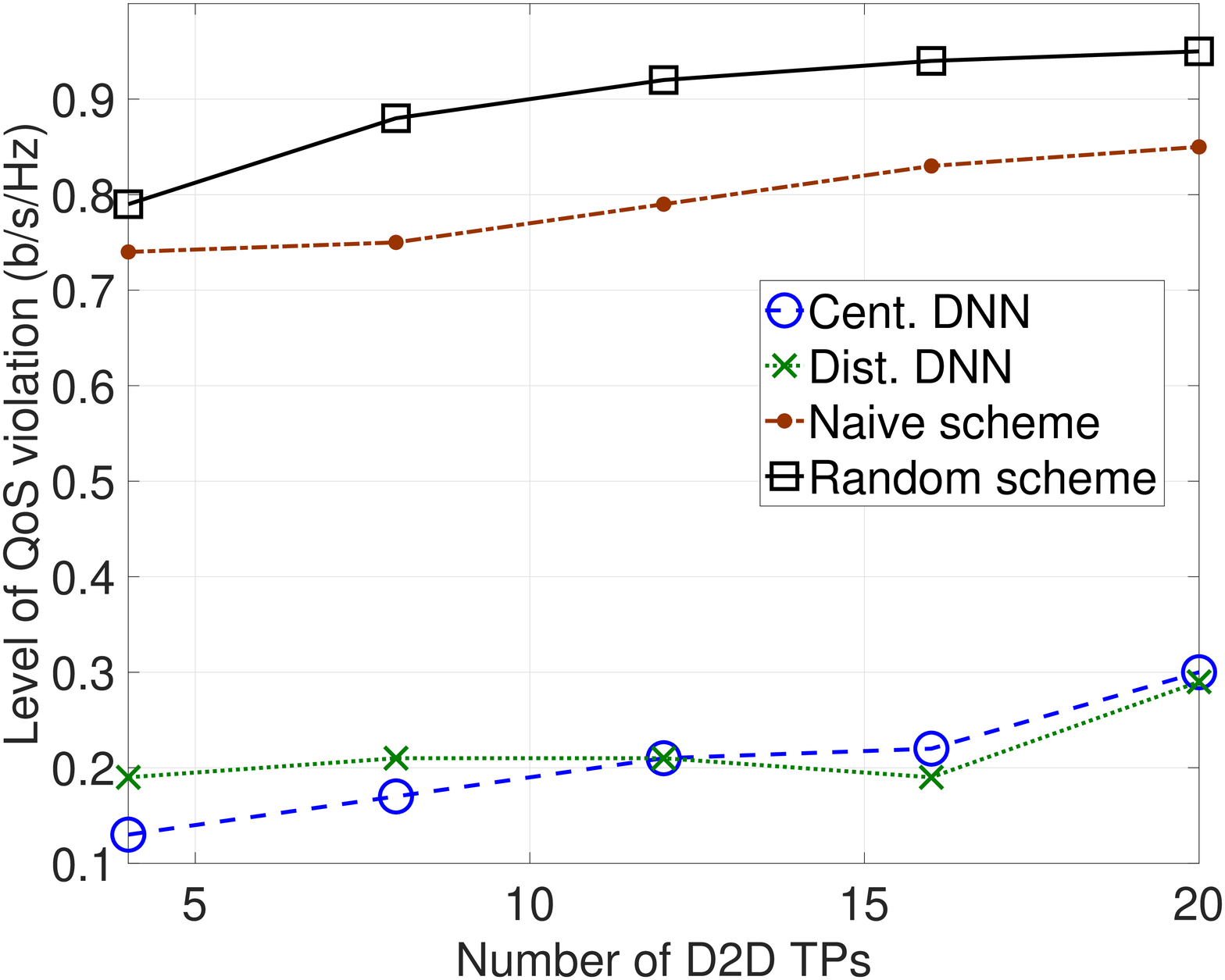}
		\label{SE_DUE_fig_c}
	}
	\end{center}\vspace{-2mm}
	\caption{Performance of proposed scheme as a function of $N$.}\vspace{-5mm}
	\label{SE_DUE_fig}
\end{figure*}

In Figs. \ref{SE_DUE_fig_a} - \ref{SE_DUE_fig_c}, we show the average SE of the
D2D TPs, the probability of QoS constraint violation, and the level of QoS violation
as a function of the number of D2D TPs, $N$. Because of the excessive computation 
time needed to determine the performance of the optimal scheme for large $N$, 
the performance of the optimal scheme is not shown. As can be seen from Fig. \ref{SE_DUE_fig_a}, the average SE 
of the D2D TPs decreases as $N$ increases due to the increased interference 
among the D2D TPs. Moreover, the naive scheme and the random 
scheme yield worse performance compared to the proposed scheme in terms of the 
average SE. In fact, the average SE of the random scheme approaches zero
as the number of D2D TPs increases due to the high probability of QoS violation, 
as confirmed by Fig. \ref{SE_DUE_fig_b}. Furthermore, Fig. \ref{SE_DUE_fig_c} shows
that the level of QoS violation of the proposed scheme increases as 
$N$ increases due to the fixed size of the DNN. Specifically, since we 
have used a relatively small DNN structure for both the centralized and distributed DNN 
schemes, the ability of the DNNs to approximate arbitrary functions is not high.
However, the dimensionality of the resource allocation strategy increases
for larger $N$, such that the performance of the proposed scheme 
deteriorates in this case.

Next, in Fig. \ref{fig_comp_time}, we compare the computation time\footnote{For these simulation results, only
the computation time for the inference phase is measured because the training of the 
DNN can be performed off-line before its actual usage.} needed for
resource allocation as a function of $N$, where the computation time 
is measured using an Intel Core i7-7700k running at 4.2 GHz with 64 GB of
memory, and the simulation codes are implemented using Python 3.7 and
Tensorflow 2.0.0. For a fair comparison, parallel 
computation, which can further reduce the computation time of DNN-based schemes,
is not employed. From the results in Fig. \ref{fig_comp_time}, we observe 
that the computation time of the optimal scheme increases exponentially with $N$,
and when $N \geq 8$, the computation time to find the optimal
resource allocation policy for one channel sample exceeds 25.9 hours, 
which is far too high for practical systems. On the other hand, the 
computation time of the proposed scheme does not increase significantly
with $N$. For $N = 100$, the computation
times for the centralized and distributed DNN schemes are 
equal to 1.6 and 24 milliseconds (not shown in Fig. \ref{fig_comp_time}), 
respectively. This suggests that the proposed scheme is applicable even for large
number of users. Furthermore, we observe that the computation time of the 
distributed DNN scheme is larger than that of the centralized DNN scheme due to the
additional binarization of the CSI.


\begin{figure}
\centerline{\includegraphics[width=9.0cm]{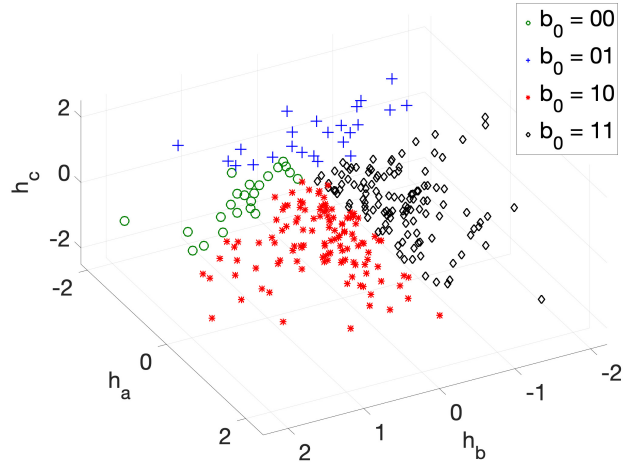}}\vspace{-2mm}
 \caption{Value of $\bm{b}_{0}$ when $B_{B}$ = 2 bits.}
\label{fig_consta}\vspace{-6mm}
\end{figure}
To illustrate the encoding of the CSI, the value of $\bm{b}_{0}$ as a function of the normalized channel gain 
is shown in Fig. \ref{fig_consta} where $\bm{b}_{0}$ is encoded by 2 bits for simplicity of presentation. 
In this simulation, the number of D2D TPs and the number of channels are assumed 
to be 2 and 1, respectively, and the normalized channel gains $\hat{h}^{1}_{0, 0}$, $\hat{h}^{1}_{1, 0}$, 
and $\hat{h}^{1}_{2, 0}$ are denoted as $h_a$, $h_b$ and $h_c$, respectively, in 
order to simplify the interpretation of the results. Note that $h_a$ corresponds to the 
channel gain of the CUE and $h_b$ and $h_c$ correspond to the gains of the interfering 
channels such that the rate of the CUE is likely large when $h_a$ is high and $h_b$ and $h_c$ 
are small. As can be observed from Fig. \ref{fig_consta}, the value of $\bm{b}_{0}$
is carefully designed by the DNN to reflect the channel conditions
such that a given $\bm{b}_{0}$ represents similar channel conditions.
For example, when the channel gain of the CUE is small and the interference
from D2D TP 1 is large, $\bm{b}_{0}$ is likely to be encoded to $00$.
This result verifies the efficiency of the proposed DNN-based approach for CSI representation.

\begin{figure}
\centerline{\includegraphics[width=7.5cm]{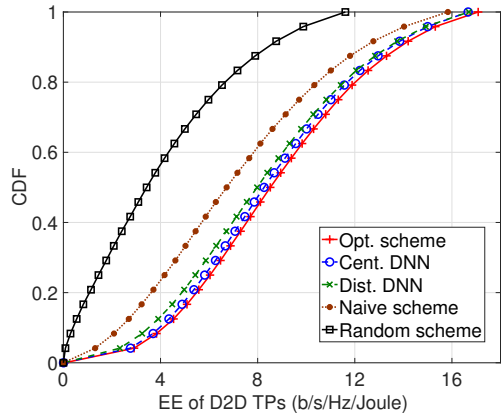}}\vspace{-2mm}
 \caption{CDF of EE for D2D TPs when $\textrm{SE}_{\textrm{thr}}$ = 0 b/s/Hz.}
\label{fig_EE_DUE_SE}\vspace{-6mm}
\end{figure}
Finally, in Fig. \ref{fig_EE_DUE_SE}, we show the performance of the 
proposed scheme for maximizing the overall EE of the D2D TPs, i.e., 
$\sum \limits_{i \in \mathbb{I} } \sum \limits_{k \in \mathbb{K} } \textrm{EE}^{k}_{i}$,
where $\textrm{EE}^{k}_{i} =  \frac{\textrm{SE}^{k}_{i}}{p_{i} + \textrm{P}_{\textrm{CIR}}}$. 
In the simulation, we assume that the circuit power of the D2D TPs, $\textrm{P}_{\textrm{CIR}}$,  is 
500 mW and $\textrm{SE}_{\textrm{thr}}$ = 0 b/s/Hz \cite{Wu2016}. 
To maximize the overall EE, we replaced the SE in the loss functions 
of the proposed schemes by the corresponding EE. From Fig. \ref{fig_EE_DUE_SE}, 
we observe that the proposed scheme also achieves a close-to-optimal performance
in this case and outperforms the conventional non-optimal schemes. 
The simulation results in Fig. \ref{fig_EE_DUE_SE} confirm the versatility of the proposed
scheme in coping with different design objectives.

\vspace{-2mm}
\section{Conclusions}\vspace{-1mm}

In this paper, we studied the resource allocation algorithm design 
for multi-channel cellular networks with D2D communication using 
a DL framework. We focused on the maximization of the overall SE 
of the D2D TPs while guaranteeing a minimum data rate for the legacy 
cellular users. Our problem formulation takes into account discrete 
optimization variables, namely channel selection and
discrete transmit power levels. In addition, we considered both 
a centralized and a distributed approach for resource allocation. In the latter case, 
encoded CSI is exchanged between the D2D users and the BS via limited 
feedback channels. We designed centralized and distributed
DNN models that approximate both the optimal resource allocation 
strategy and the optimal encoding strategy of the local CSI for a limited
feedback capacity. In order to facilitate the training of the proposed DNN 
models, a hybrid supervised and unsupervised learning strategy with novel 
loss functions was devised and shown to enable efficient training based on a
few ground-truth labels. Through computer simulations, 
we confirmed that the proposed resource allocation schemes achieve near-optimal 
performance with low computation time, which underlines their applicability 
in practical systems. Specifically, we found that the distributed resource 
allocation scheme achieves a similar performance as the centralized
scheme, which validates the effectiveness of the proposed CSI representation
using a DNN. Furthermore,  we verified the scalability of the proposed schemes
for a large number of users in view of computation time and 
showed that the proposed DNN model can be applied to achieve various design 
objectives, e.g., EE maximization.

\vspace{-2mm}
\bibliographystyle{IEEEtran}
\bibliography{IEEEabrv,mybibfile}

\end{document}